\documentclass[12pt, draftclsnofoot, onecolumn]{IEEEtran}

\usepackage{amsmath}
\usepackage{amssymb}
\usepackage{amsfonts}
\usepackage{graphicx}
\usepackage{epsfig}
\usepackage{subfigure}
\usepackage{psfrag}
\usepackage{cite}
\usepackage{latexsym}
\usepackage{url}
\usepackage{color}
\usepackage{multirow}
\usepackage{mathtools}
\usepackage{bm}
\usepackage{algorithm}
\usepackage{algorithmic}
\usepackage{verbatim}
\usepackage{indentfirst}
\usepackage{hyperref}
\PassOptionsToPackage{bookmarks={false}}{hyperref}

\IEEEoverridecommandlockouts
\newtheorem{proposition}{\underline{Proposition}}[section]

\begin{document}
\title{Joint Maneuver and Beamforming Design for UAV-Enabled Integrated Sensing and Communication}
\author{Zhonghao Lyu, Guangxu Zhu, and Jie Xu\\
\thanks{Part of this paper has been submitted to IEEE International Conference on Communications (ICC) 2022 \cite{ZL2022}.}
\thanks{Z. Lyu and J. Xu are with the Future Network of Intelligence Institute (FNii) and the School of Science and Engineering (SSE), The Chinese University of Hong Kong (Shenzhen), Shenzhen, China (e-mail: zhonghaolyu@link.cuhk.edu.cn, xujie@cuhk.edu.cn). J. Xu is the corresponding author. }
\thanks{
G. Zhu is with Shenzhen Research Institute of Big Data, Shenzhen, China  (e-mail: gxzhu@sribd.cn). }
}
\maketitle

\begin{abstract}
This paper studies the unmanned aerial vehicle (UAV)-enabled integrated sensing and communication (ISAC), in which UAVs are dispatched as aerial dual-functional access points (APs) that can exploit the UAV maneuver control and strong line-of-sight (LoS) aerial-to-ground (A2G) links for efficient ISAC. In particular, we consider a scenario with one UAV-AP equipped with a vertically placed uniform linear array (ULA), which sends combined information and sensing signals to communicate with multiple users and at the same time sense potential targets at interested areas on the ground. Our objective is to jointly design the UAV maneuver together with the transmit beamforming for optimizing the communication performance while ensuring the sensing requirements. First, we consider the quasi-stationary UAV scenario, in which the UAV is deployed at an optimizable location over the whole ISAC mission period. In this case, we jointly optimize the UAV deployment location, as well as the transmit information and sensing beamforming to maximize the weighted sum-rate throughput of communication users, subject to the sensing beampattern gain requirements and transmit power constraint at the UAV. Although the joint UAV deployment and beamforming problem is non-convex, we find a high-quality solution by using the techniques of successive convex approximation (SCA) and semidefinite relaxation (SDR), together with a two-dimensional (2D) location search. Next, we consider the fully mobile UAV scenario, in which the UAV can fly over different locations during the ISAC mission period. In this case, we optimize the UAV flight trajectory, jointly with the transmit beamforming over time, to maximize the average weighted sum-rate throughput of communication users over the whole period, subject to the sensing beampattern gain requirements and transmit power constraints over different time slots, as well as practical flight constraints. While the joint UAV trajectory and beamforming problem is more challenging to solve, we propose an efficient algorithm by adopting the alternating optimization together with SCA. Finally, numerical results are provided to validate the superiority of our proposed designs as compared to various benchmark schemes with heuristic maneuver designs.
\end{abstract}


\section{Introduction}
Recent advancements in 5G-and-beyond networks are envisioned to enable many environment- and location-aware intelligent applications such as auto-driving, remote healthcare, and smart industry. To support these applications, 5G-and-beyond networks are expected to provide high-precision sensing capabilities, in addition to conventional wireless communication services \cite{ML2020}. Towards this end, integrated sensing and communication (ISAC) \cite{FL2021,FL2020} (a.k.a. radar-communica-tion (RadCom) \cite{WY2021}, dual-functional radar communication (DFRC) \cite{XL2020,CX2021}, joint communication and radar sensing (JCAS) \cite{ML2020}) has recently been recognized as one of the key technologies towards 5G-and-beyond wireless networks, and thus attracted tremendous research interests from both academia and industry \cite{ML2020,FL2021,FL2020,WY2021,Eri,Int,FLCM2018,XL2020,CX2021,HH2021,YL2019}. ISAC provides various advantages over conventional wireless networks with communication functionality only or with coexisting radar and communication that are separately designed  \cite{RS2012,JA2017,SS2012}. First, ISAC allows wireless infrastructures and scarce spectrum resources to be seamlessly shared for the dual use of both sensing and communication, thus leading to significantly enhanced spectrum, energy, and hardware utilization efficiency. Next, with the on-going deployment of millimeter wave and massive multiple-input-multiple-output (MIMO), the communication signals are very promising to provide ultra-high sensing accuracy and resolution \cite{YW2019}. Furthermore, the integrated sensing functionality can also benefit the communication design \cite{FL2021}, e.g., the sensory data can be leveraged to facilitate the beam training in vehicle-to-everything (V2X) networks \cite{WY2021}.

The use of multiple antennas or MIMO has played an important role in the separate development of sensing and communication (see, e.g., \cite{JL2007,RW2018} and the references therein). Motivated by them, the MIMO ISAC with transmit beamforming has also attracted growing research interests recently (e.g., \cite{FL2020,FL2021,WY2021,FLCM2018,XL2020,CX2021,HH2021,YL2019}). For instance, the authors in \cite{FLCM2018} considered to reuse the communication signals for sensing, in which the information beamforming is designed to minimize the sensing beampattern matching error, while ensuring the communication performance in terms of the minimally required signal-to-interference-plus-noise ratio (SINR) at each user. Besides reusing information signals for ISAC, the authors in \cite{XL2020} and \cite{HH2021} proposed to additionally send dedicated sensing signals, which are able to provide more degrees of freedom (DoF) for sensing, thus leading to further enhanced communication and sensing performances. However, these prior works mainly focused on the MIMO ISAC in terrestrial networks, which faces inherent limitations, especially for sensing. First, the target detection and parameters estimation in sensing generally depend on the explicit line-of-sight (LoS) links between sensing transceivers (e.g., ISAC base station (BSs) or access points (APs)) and targets. However, there normally exist many surrounding obstacles and scatters on the ground, which may block the LoS links and create many non-LoS (NLoS) signal paths or clutters, thus making sensing difficult or even infeasible. Next, large signal power is generally required to achieve high-precision sensing at long distances, but the terrestrial BSs/APs are generally with limited transmit power. In this case, if the targets are located far away from BSs/APs, then the sensing performance may seriously degrade due to the severe round-trip propagation power loss of the echoed signals. 

Driven by the recent development of unmanned aerial vehicle (UAV)-enabled wireless communications \cite{QWJXYZ2021}, we expect that UAVs can also be utilized as a new type of aerial ISAC platforms to relieve the above limitations, especially for emergence scenarios after disaster or at temporary hot spot areas in outdoor. First, due to the relatively high altitude of UAVs, it is highly likely that there exist LoS connection for the aerial-to-ground (A2G) links \cite{MM2019,WK2019,AA2018}, which is thus a natural fit for sensing. On the other hand, due to the fully-controllable mobility, UAVs can either be deployed as quasi-stationary APs or dispatched as fully mobile APs close to the interested areas with targets, thus relieving the high-power requirements for sensing. In the literature, there have been various prior works investigating UAV-enabled wireless communications under different setups, such as relaying \cite{YZRZ2016,JC2017,HW2018}, interfering networks \cite{QWYZ2018,RH2020}, multiple access \cite{PL2020} and broadcast channels \cite{QWJXZR2018}, and energy efficient communications \cite{YZRZ2017,SS2021,TZ2020}. Along this direction, the joint UAV trajectory and communications design is critical to enhance the communication performance. Moreover, the multi-antenna or MIMO techniques have also been integrated in UAV-enabled communication systems to enhance the system spectrum efficiency and network coverage by exploiting the spatial multiplexing and beamforming gains \cite{ZX2021}. In this case, the UAV's deployment and trajectory designs should be  optimized jointly with the transmit beamforming \cite{ZXHD2020,JZ2020,LLSZ2019,DX2020}. On the other hand, there has been another line of research studying the UAV-enabled wireless sensing \cite{HE2012,AG2015,ZS2016,FK2017}. For instance, the UAV synthetic aperture radar (SAR) has been widely implemented nowadays, in which the Doppler shift arising from the relative movement between the UAV and targets is exploited to improve the azimuth resolution (e.g., \cite{HE2012,AG2015,ZS2016}). Despite the above separate research progress in UAV-enabled communication and UAV-enabled sensing, however, the research on UAV-enabled ISAC is still in its infancy stage. In particular, how to jointly design the ISAC and the deployment location or flight trajectory of the UAV to enhance sensing and communication performances is an appealing yet challenging problem, especially when the UAV is equipped with multiple antennas with LoS A2G channels. This thus motivates our work in this paper.

\begin{figure}[h]
\centering
 \epsfxsize=1\linewidth
    \includegraphics[width=9cm]{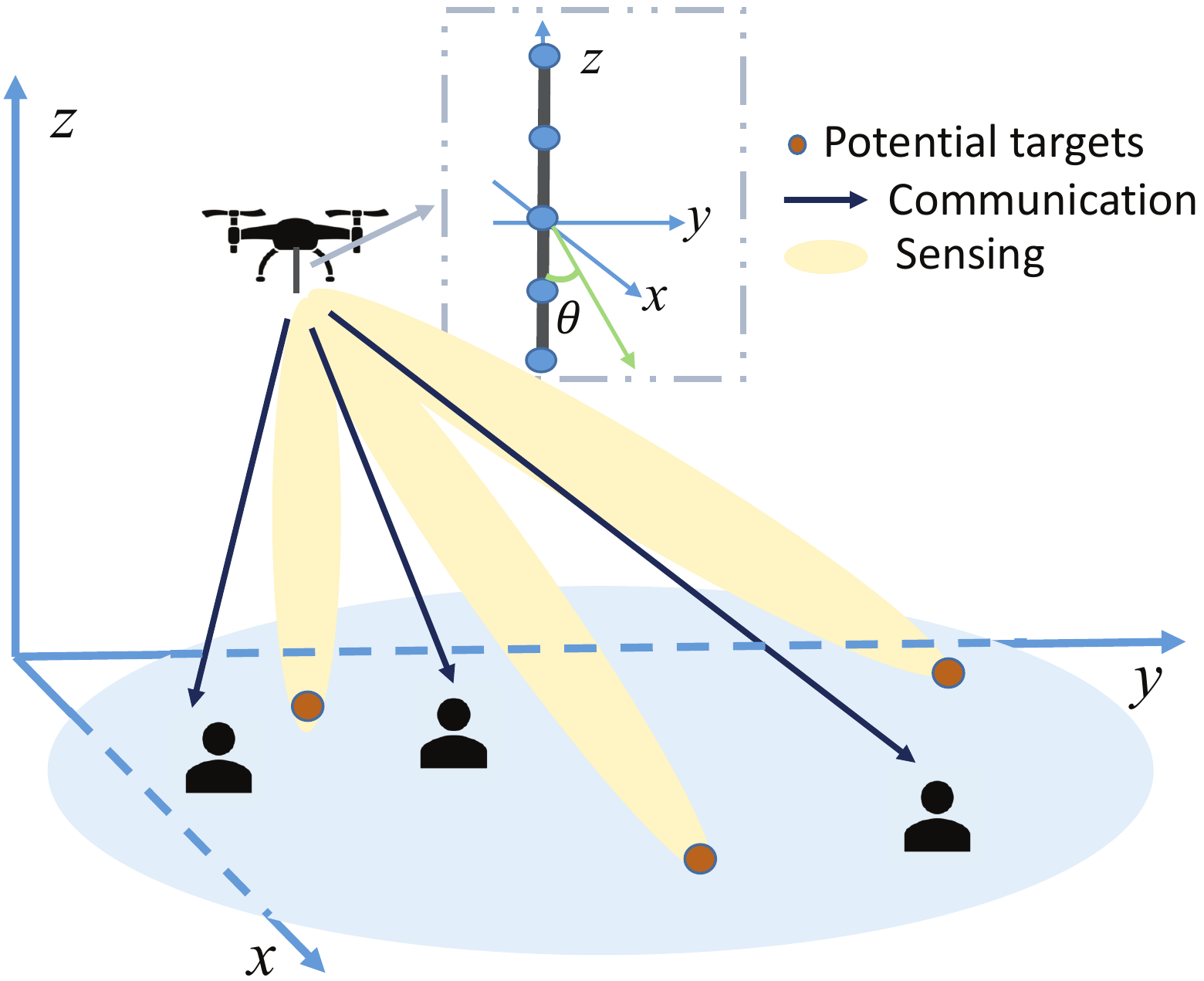}
\vspace{-4pt}
\caption{Illustration of the UAV-enabled ISAC system with a ULA vertically placed at the UAV.}
\end{figure}

This paper studies a UAV-enabled ISAC system as shown in Fig. 1, in which a UAV equipped with a vertically placed uniform linear array (ULA) is employed as an aerial dual-functional AP for efficient ISAC. Particularly, the UAV sends unified information and sensing signals to communicate with multiple users and perform radar sensing towards potential targets at interested areas simultaneously. Under this setup, we aim to design the UAV maneuver jointly with the transmit beamforming for optimizing the communication performance while guaranteeing the radar sensing requirements. The main results of the paper are summarized as follows.
\begin{itemize}
    \item First, we consider the quasi-stationary UAV scenario, where the UAV is fixed at an optimizable location to perform ISAC over the whole mission period. We aim at maximizing the weighted sum-rate throughput of the users via jointly optimizing the UAV’s deployment location, as well as the transmit information and sensing beamforming, subject to the transmit power constraint and sensing beampattern gain requirements. Although the above problem is non-convex and difficult to be optimally solved in general, we propose to find a high-quality solution via using the techniques of successive convex approximation (SCA) \cite{QT2010} and semidefinite relaxation (SDR) \cite{ZL2014}, together with a two-dimensional (2D) location search.
    \item Next, we consider the fully mobile UAV scenario, where the UAV flies from the predetermined initial to final locations over the ISAC mission period. In this case, we design the UAV's trajectory, jointly with the information and sensing beamforming over time to maximize the average weighted sum-rate throughput of communication users, subject to the sensing beampattern gain and transmit power constraints over time, as well as practical flight constraints. However, the joint UAV trajectory and beamforming problem is even more challenging to solve due to the involvement of more UAV trajectory variables. To tackle this issue, we propose an efficient algorithm by adopting the alternating optimization together with SCA.
    \item Finally, numerical results are presented to validate the performance of our proposed designs versus benchmark schemes with heuristic maneuver designs. It is shown that when the sensing beampattern gain threshold is high, then the UAV should be deployed close to or fly towards the sensing area. By contrast, if the sensing beampattern gain threshold is low, then the UAV is deployed or flies close to an optimized location that is close to but not exactly above the users. This shows that the UAV's deployment and trajectory design plays an important role in balancing the inherent sensing-communication performance tradeoff. 
\end{itemize}

The remainder of this paper is organized as follows. Section II introduces the system model of UAV-enabled ISAC system. Section III formulates the sensing-constrained weighted sum rate maximization problems of interest, and checks the feasibility of the formulated problems. Sections IV and V address the weighted sum rate maximization problems for two scenarios with quasi-stationary and mobile UAVs, respectively. Section VI provides numerical results to demonstrate the efficiency of our proposed designs. Section VII concludes this paper.

{\emph {Notations}}: Boldface letters refer to vectors (lower case) or matrices (upper case). For an arbitrary-size matrix ${\boldsymbol A}$, ${\rm rank}({\boldsymbol A})$, ${\boldsymbol A}^{\rm H}$, ${\boldsymbol A}^{\rm T}$, and $[\boldsymbol{A}]_{p,q}$ denote its rank, conjugate transpose, transpose, and the element in the $p$-th row and $q$-th column, respectively. For a square
matrix ${\boldsymbol B}$, ${\rm tr}({\boldsymbol B})$ and ${\boldsymbol B}^{-1}$ denote its trace and inverse, respectively, and ${\boldsymbol B}\succeq {\boldsymbol 0}$ means that ${\boldsymbol B}$ is positive semidefinite. ${\boldsymbol I}$ denotes an identity matrix, and ${\boldsymbol 0}$ denotes an all-zero matrix. $\mathbb{C}^{M \times N}$ denotes the space of ${M \times N}$ complex matrices. $\mathbb{E}( \cdot )$ denotes the statistical expectation. $\|\cdot\|$ denotes the Euclidean norm of a complex vector, and $|\cdot|$ denotes the magnitude of a complex number.

\section{System Model}
We consider a UAV-enabled ISAC system as shown in Fig. 1, in which a UAV is dispatched as an aerial dual-functional AP to perform downlink communication with $K \ge 1$ ground users and radar sensing towards potential ground targets at the same time. Let $\mathcal{K} \triangleq \{ 1,...,K\}$ denote the set of ground communication users. It is assumed that each communication user is equipped with one single receive antenna, and the UAV is equipped with a ULA with $M$ antennas that are placed vertically to the horizontal plane, similarly as in \cite{XYYH2021}. Notice that the vertical ULA placement is considered at the UAV to facilitate the technical derivation, since in this case the UAV's orientation is irrelevant to the angle of departure (AoD) of the radiated communication and sensing signal beams, thus simplifying the UAV's trajectory design\footnote{Notice that under other antenna configurations (e.g.,  horizontally placed uniform planar array (UPA)), the UAV's orientation will become an additional design DoF for improving the performance of ISAC systems, the details of which, however, will be left for future work.}.  

We consider a finite ISAC mission period $\mathcal{T} \triangleq [0,T]$ with duration $T > 0$, which is discretized into $N$ time slots each with duration ${\varDelta _t} = T/N$. Let $\mathcal N \triangleq \{1,...N\}$ denote the set of time slots. Here, ${\varDelta _t}$ is chosen to be sufficiently small, during which the UAV's location is assumed to be approximately unchanged to facilitate the trajectory and beamforming design for ISAC. We consider a three-dimensional (3D) Cartesian coordinate system, where the location of each user $k\in \mathcal{K}$ is fixed at ${(x_k,y_k,0)}$ with ${{\boldsymbol {u}}_k}= {(x_k,y_k)}$ denoting its horizontal location. Let $(x[n],y[n],H)$ denote the time-varying location of the UAV at time slot $n\in\mathcal N$, where  ${{\boldsymbol{q}}}[n]={(x[n],y[n])} $ denotes its horizontal location, and $H$ denotes its flying altitude that is fixed over time depending on certain flight regulations.

We consider that the UAV transmits information signals for users, together with dedicated sensing signals that can provide additional design DoFs for facilitating the sensing \cite{XL2020,HH2021,FLCM2018}. In particular, consider one particular time slot $n\in \mathcal N$. Let ${s_k}[n]$ denote the desired information signal by user $k \in \mathcal{K}$, ${\boldsymbol{w}}_{k}[n] \in \mathbb{C}^{M \times 1}$ denote the corresponding transmit beamforming vector, and ${{\boldsymbol{s}}_0}[n] \in \mathbb{C}^{M \times 1}$ denote the dedicated radar sensing signal at slot $n$. It is assumed that the communication signals ${\{{s_k}[n]\}^K_{k=1}}$ are independent circularly symmetric complex Gaussian (CSCG) random variables with zero mean and unit variance, i.e., $s_k[n] \sim \mathcal{CN}(0,1)$, and the dedicated sensing signal ${\boldsymbol s}_0[n]$ is an independently generated random vector with zero mean and covariance matrix ${{\boldsymbol{R}}_{ s}}[n] = \mathbb{E}({{\boldsymbol{s}}_0}[n]{\boldsymbol{s}}_0^{\rm{H}}[n]) \succeq \boldsymbol{0}$.
Accordingly, the transmitted signal by the UAV at time slot $n$ is given by
\begin{align}\label{signal x}
{\boldsymbol{x}}[n] = \sum\limits_{k = 1}^K {{{\boldsymbol{w}}_k}[n]{s_k}[n] + }{{\boldsymbol{s}}_0}[n],\forall n \in \mathcal{N}.
\end{align}
The average transmit power by the UAV at time slot $n$ is denoted as $\mathbb{E}(\|\boldsymbol{x}[n]\|^2) =\sum\nolimits_{k = 1}^K {\|{{\boldsymbol{w}}_k}[n]\|{^2}} \\ + {\rm{tr}}({{\boldsymbol{R}}_{ s}}[n])$. Suppose that the maximum transmit power at the UAV is $P_{\rm max}$. We thus have the transmit power constraints as
\begin{align}\label{transmit power}
\sum\nolimits_{k = 1}^K {\|{{\boldsymbol{w}}_k}[n]\|{^2}}  + {\rm{tr}}({{\boldsymbol{R}}_{ s}}[n]) \le {P_{\max }}, \forall n\in \mathcal N.
\end{align}

First, we consider the information reception at the $K$ users. Due to the relatively high altitude of the UAV, there generally exists a strong LoS link between the UAV and each ground user. As such, in this paper we consider the LoS channel model\footnote{Notice that the LoS channel model is considered for gaining the essential design insights. This model is consistent with various measurement results \cite{WK2019,AA2018}, and is widely adopted in previous works (e.g., \cite{QWYZ2018,YZRZ2017,QWJXZR2018}). Our design principle can also be extended to scenarios with other channel models with additional NLoS links (e.g., Rician fading). }, based on which the channel vector from the UAV to user $k$ at time slot $n$ is denoted as 
\begin{align}\label{channel}
{\boldsymbol{h}}_k({\boldsymbol{{q}}}[n],{{\boldsymbol{u}}_k})= \sqrt {\beta {d^{ - 2 }({\boldsymbol q}[n],{{\boldsymbol u}_k})}} {\boldsymbol{a}}({\boldsymbol{{q}}}[n],{{\boldsymbol{u}}_k})=\sqrt {\frac{\beta }{{{H^2} + \|{\boldsymbol{q}}[n] - {{\boldsymbol{u}}_k}\|{^2}{\rm{ }}}}} {\boldsymbol{a}}({\boldsymbol{q}}[n],{{\boldsymbol{u}}_k}),
\end{align}
where $\beta$ denotes the channel power gain at a reference distance $d_0=1~{\rm m}$, $d({\boldsymbol q}[n],{{\boldsymbol u}_k})=\sqrt {{{{H^2} + \|{\boldsymbol{q}}[n] - {{\boldsymbol{u}}_k}\|{^2}}}}$ denotes the distance between the UAV and user $k$ at slot $n$, and ${\boldsymbol{a}}({\boldsymbol{q}}[n],{{\boldsymbol{u}}_k})$ denotes the steering vector towards user $k$. More specifically, the steering vector ${{\boldsymbol{a}}}({\boldsymbol{{q}}}[n],{{\boldsymbol{u}}_k})$ is expressed as
\begin{align}\label{steering vector}
{\boldsymbol{a}}({\boldsymbol{{q}}}[n],{{\boldsymbol{u}}_k}) = {\left[ {1,{e^{j2\pi \frac{d}{\lambda }\cos \theta({\boldsymbol{{q}}}[n],{{\boldsymbol{u}}_k}) }}, \ldots ,{e^{j2\pi \frac{d}{\lambda }(M - 1)\cos \theta({\boldsymbol{{q}}}[n],{{\boldsymbol{u}}_k}) }}} \right]^{\rm T}},
\end{align}
where $\lambda$ and $d$ denote the carrier wavelength and the spacing between two adjacent antennas, respectively, and $\theta({\boldsymbol{{q}}}[n],{{\boldsymbol{u}}_k})$ denotes the AoD corresponding to user $k$ with\footnote{It follows from \eqref{theta} that the ground users/locations lying in the same circle centered at the UAV's horizontal location will experience the same steering vector.}
\begin{align}\label{theta}
\theta ({\boldsymbol{{q}}}[n],{{\boldsymbol{u}}_k}) = {\rm{arccos}}\frac{H}{{\sqrt {\|{\boldsymbol{q}}[n] - {{\boldsymbol{u}}_k}\|{^2}{\rm{ + }}{H^2}} }}.
\end{align}

As a result, the received signal by user $k$ at slot $n$ is given as
\begin{align}\label{received signal}
{z_k}[n] = {\boldsymbol{h}}_k^{\rm H}({\boldsymbol{{q}}}[n],{{\boldsymbol{u}}_k}){\boldsymbol{x}}[n] + {v_k}[n] = {\boldsymbol{h}}_k^{\rm H}({\boldsymbol{{q}}}[n],{{\boldsymbol{u}}_k})\left( \sum\limits_{k = 1}^K {{{\boldsymbol{w}}_k}[n]{s_k}[n] + } {{\boldsymbol{s}}_0}[n]\right) + {v_k}[n],
\end{align}
where $v_k[n]$ denotes the additive white Gaussian noise (AWGN) at the user receiver that is a CSCG random variable with zero mean and variance $\sigma _k^2$.

It is observed from \eqref{received signal} that each ground user $k$ suffers from not only the co-channel interference induced by other users' desired information signals ${\{ {s_i}[n]\} _{i \ne k}}$, but also that by the dedicated radar signal ${{\boldsymbol{s}}_0}[n]$. Therefore, the received SINR by user $k$ at time slot $n$ is
\begin{align}\label{SINR}
{\gamma _k}[n]= \frac{{{{\left| {{\boldsymbol{h}}_k^{\rm H}({\boldsymbol{{q}}}[n],{{\boldsymbol{u}}_k}){{\boldsymbol{w}}_k}}[n] \right|}^2}}}{{\sum\nolimits_{i=1,i \ne k}^K {{{\left| {{\boldsymbol{h}}_k^{\rm H}({\boldsymbol{{q}}}[n],{{\boldsymbol{u}}_k}){{\boldsymbol{w}}_i}}[n] \right|}^2}}  + {\boldsymbol{h}}_k^{\rm H}({\boldsymbol{{q}}}[n],{{\boldsymbol{u}}_k}){{\boldsymbol{R}}_s}[n]{\boldsymbol{h}}_k({\boldsymbol{{q}}}[n],{{\boldsymbol{u}}_k}) + \sigma _k^2}}.
\end{align}
Under the Gaussian signalling, the achievable rate by user $k$ at slot $n$ (in bits-per-second-per-Hertz (bps/Hz)) is 
\begin{align}\label{rate}
r_k[n] = {\log _2}(1+{\gamma _k}[n]).
\end{align}

Next, we consider the radar sensing towards potential targets at the interested area. In particular, suppose that the UAV aims to sense potential targets at a finite number of $J$ locations on the ground, whose horizontal locations are denoted by ${{\boldsymbol{m}}}_j$'s, $j\in {\mathcal J}\triangleq \{1,\ldots, J\}$. Here, the values of ${\boldsymbol m}_j$'s are predetermined based on the specific sensing tasks at the UAV. For instance, if the UAV performs the detection task without knowing the presence and thus locations of potential targets, then ${\boldsymbol m}_j$'s can be set as uniformly sampled positions over the whole area of interest. By contrast, if the UAV performs the tracking task with the targets' locations roughly known {\it a-priori}, then ${\boldsymbol m}_j$'s can be set as the possible locations of these targets for facilitating the tracking. In general, a larger value of $J$ may be needed for more accurate sensing, but at the cost of higher computational complexity. Notice that similarly as in prior works (e.g., \cite{XL2020,HH2021,FLCM2018}), we consider that the communication signals ${\{{s_k}[n]\}^K_{k=1}}$ and the dedicated sensing signal ${\boldsymbol s}_0[n]$ are jointly exploited for sensing. In this case, the transmit beampattern gain towards location $\boldsymbol{m}_j$ is
\begin{align}\label{beampattern gain}
&\zeta({\boldsymbol{{q}}}[n],{{\boldsymbol{m}}_j}) ={\mathbb{E}}[|{{\boldsymbol{a}}^{\rm H}}({\boldsymbol{{q}}}[n],{{\boldsymbol{m}}_j}){\boldsymbol{x}}[n]|^2] \nonumber \\
=& {{\boldsymbol{a}}^{\rm H}}({\boldsymbol{{q}}}[n],{{\boldsymbol{m}}_j})\left(\sum\nolimits_{k = 1}^K {{{\boldsymbol{w}}_k}[n]{\boldsymbol{w}}_k^H[n] + {{\boldsymbol{R}}_{s}}}[n] \right){{\boldsymbol{a}}}({\boldsymbol{{q}}}[n],{{\boldsymbol{m}}_j}),
\end{align}
where ${\boldsymbol{a}}({\boldsymbol{{q}}}[n],{{\boldsymbol{m}}_j})$ denotes the steering vector as defined in \eqref{steering vector}.

\section{Problem Formulation and Feasibility Checking}
\subsection{Problem Formulation}
Our objective is to maximize the communication rate performance at multiple ground users while ensuing the sensing requirements by considering two scenarios with quasi-stationary and mobile UAVs, in which the UAV stays at one single optimized location over the whole mission period and can freely move from one location to another, respectively. 

First, we consider the quasi-stationary UAV scenario, in which the UAV is fixed at an optimizable location ${\boldsymbol{q}} = ({x^{\rm q}},{y^{\rm q}})$ over the whole mission duration, i.e., ${{\boldsymbol q}[n] = {\boldsymbol q}}, \forall n\in\mathcal {N}$. In this scenario, the time index $n$ is discarded for notational convenience. Our objective is to maximize the weighted sum rate $\sum\nolimits_{k = 1}^K \alpha_k {r_k}$ with ${r_k}$ given in (8), while guaranteeing the sensing performance at a given set of locations ${{{{\boldsymbol{m}}}_1},{{{\boldsymbol{m}}}_2}, \ldots ,{{{\boldsymbol{m}}}_J}}$, by jointly optimizing the information and sensing beamforming vectors $\{{{\boldsymbol{w}}_k}\}$ and ${{\boldsymbol{R}}_{ s}}$ as well as the UAV's deployment location ${\boldsymbol q}$. Here, $\alpha_k$ denotes the weight of each user $k$, where a larger value of $\alpha_k$ means that user $k$ has a higher priority in rate maximization. In order to properly illuminate potential targets at interested areas, the transmit beampattern gain at interested sensing locations ${\boldsymbol m}_j$ should be no less than a certain threshold that is proportional to the square of the UAV's distance with that location ${{d({\boldsymbol q},{{\boldsymbol m}_j})}}$ (i.e., ${{d^{2 }({\boldsymbol q},{{\boldsymbol m}_j})}}\Gamma$ with $\Gamma$ being a predetermined threshold). As a result, the sensing constrained weighted sum rate maximization problem in the quasi-stationary UAV scenario is formulated as 
\begin{subequations}\label{P1}
\begin{align}
\text{(P1)}:\mathop {\max }\limits_{\{{{\boldsymbol{w}}_k}\},\atop {{\boldsymbol{R}}_{ s}}\succeq \boldsymbol{0},{\boldsymbol{q}}} ~&  {\sum\nolimits_{k = 1}^K  \alpha_k{{\log _2}\left(1 + \frac{{{{\left| {{\boldsymbol{h}}_k^{\rm{H}}({\boldsymbol{q}},{{\boldsymbol{u}}_k}){{\boldsymbol{w}}_k}} \right|}^2}}}{{\sum\nolimits_{i=1,i \ne k}^K {{{\left| {{\boldsymbol{h}}_k^{\rm{H}}({\boldsymbol{q}},{{\boldsymbol{u}}_k}){{\boldsymbol{w}}_i}} \right|}^2}}  + {\boldsymbol{h}}_k^{\rm{H}}({\boldsymbol{q}},{{\boldsymbol{u}}_k}){{\boldsymbol{R}}_{{s}}}{\boldsymbol{h}}_k^{}({\boldsymbol{q}},{{\boldsymbol{u}}_k}) + \sigma _k^2}}\right)} }   \nonumber \\
\mathrm{s.t.}~&
{\boldsymbol{a}}^{\rm H}({\boldsymbol{q}},{{\boldsymbol{m}}_j})\left(\sum\nolimits_{k = 1}^K {{{\boldsymbol{w}}_k}{\boldsymbol{w}}_k^{\rm H} + {{\boldsymbol{R}}_{ s}}}\right ){\boldsymbol{a}}({\boldsymbol{q}},{{\boldsymbol{m}}_j}) \ge {{d^{2 }({\boldsymbol q},{{\boldsymbol m}_j})}}\Gamma ,\forall {j} \in \mathcal{J} \label{P1-a}\\
~
&\sum\nolimits_{k = 1}^K {\|{{\boldsymbol{w}}_k}\|{^2}}  + {\rm{tr}}({{\boldsymbol{R}}_{s}}) \le {P_{\max }} ,\label{P1-b}
\end{align}
\end{subequations}
where \eqref{P1-a} and \eqref{P1-b} denote the sensing beampattern gain  and the  transmit power constraint, respectively. Notice that problem (P1) is quite challenging to be optimally solved, as the objective function is non-concave and the constraints in \eqref{P1-a} is non-convex, due to the coupling between the UAV's deployment location and the transmit beamforming. We will address problem (P1) in Section IV.

Next, we consider the fully mobile UAV scenario. Suppose that ${\boldsymbol{\hat q}}^{\rm {I}} = {(x^{\rm {I}},y^{\rm {I}})}$ and ${\boldsymbol{\hat q}}^{\rm {F}} = {(x^{\rm {F}},y^{\rm {F}})}$ denote the initial and final horizontal locations of the UAV, which are pre-determined depending on the UAV mission. Let $ {\tilde V}_{\max }$ denote the maximum flight speed, and ${V_{\max }}={\tilde V_{\max }}{\Delta _t}$ denote the UAV's maximum displacement over two consecutive slots. As a result, we have the following flight constraints on the UAV:
\begin{align}\label{initial loc}
{{\boldsymbol{q}}}[1] = {\boldsymbol{\hat q}}^{\rm {I}},
\end{align}
\begin{align}\label{final loc}
{{\boldsymbol{q}}}[N] = {\boldsymbol{\hat q}}^{\rm {F}} ,
\end{align}
\begin{align}\label{speed constr}
\left\| {{{\boldsymbol{q}}}[n + 1] - {{\boldsymbol{q}}}[n]} \right\| \le V_{\max },\forall n \in  \mathcal{N}\backslash \{N\}.
\end{align}
In this scenario, our objective is to maximize the average weighted sum-rate throughput $\frac{1}{N}\sum_{n=1}^N$$\\$$\sum_{k=1}^K \alpha_k r_k[n]$, by jointly optimizing the UAV's trajectory $\{\boldsymbol{q}[n]\}$, as well as the transmit information and sensing beamforming $\{\boldsymbol{w}_k[n], \boldsymbol{R}_s[n]\}$, subject to the sensing requirements and transmit power constraints over different time slots, as well as the UAV's flight constraints in \eqref{initial loc}, \eqref{final loc}, and \eqref{speed constr}. Accordingly, the sensing-constrained weighted sum rate maximization problem via joint trajectory and beamforming optimization is formulated as 
\begin{comment}
\begin{subequations}\label{P2}
\begin{align}
\text{(P2)}:\mathop {\max }\limits_{\{{\boldsymbol w}_k[n],\atop {\boldsymbol R}_s[n]\succeq \boldsymbol{0},{\boldsymbol q}[n]\}} \!& \frac{1}{N}\sum\limits_{n = 1}^N \! {\sum\limits_{k = 1}^K \alpha_k{\log _2}\left(1\!+\! \frac{{{{\left| {{\boldsymbol{h}}_k^{\rm{H}}({\boldsymbol{q}}[n],{{\boldsymbol{u}}_k}){{\boldsymbol{w}}_k}[n]} \right|}^2}}}{{\sum\limits_{i \in \mathcal{K},i \ne k}\! {{{\left| {{\boldsymbol{h}}_k^{\rm{H}}({\boldsymbol{q}}[n],{{\boldsymbol{u}}_k}){{\boldsymbol{w}}_i}[n]} \right|}^2}}  \!+\! {\boldsymbol{h}}_k^{\rm{H}}({\boldsymbol{q}}[n],{{\boldsymbol{u}}_k}){{\boldsymbol{R}}_{{s}}}[n]{{\boldsymbol{h}}_k}({\boldsymbol{q}}[n],{{\boldsymbol{u}}_k}) \!+\! \sigma _k^2}}\right) }   \nonumber\\
\mathrm{s.t.}~&
{\boldsymbol{a}}^{\rm H}({\boldsymbol{{q}}}[n],{{\boldsymbol{m}}_j})\left(\sum\limits_{k = 1}^K {{{\boldsymbol{w}}_k}[n]{\boldsymbol{w}}_k^{\rm H}[n] + {{\boldsymbol{R}}_{ s}}} [n]\right){\boldsymbol{a}}({\boldsymbol{{q}}}[n],{{\boldsymbol{m}}_j}) \nonumber\\
&~~~~~~~~~~~~~~~~~~~~~~~~\ge d^2({\boldsymbol q}[n],{{\boldsymbol m}_j})\Gamma ,\forall j \in \mathcal J,n \in \mathcal N \label{P2-a beampattern}\\
~
&\sum\nolimits_{k = 1}^K {\|{{\boldsymbol{w}}_k}[n]\|{^2}}  + {\rm{tr}}({{\boldsymbol{R}}_{ s}}[n]) \le {P_{\max }}, \forall n\in \mathcal N  \label{P2-b tras power}\\
~
&(11),~(12),~\text{and}~(13).\nonumber
\end{align}
\end{subequations}
\end{comment}
\begin{subequations}\label{P2}
\begin{align}
\text{(P2)}:\mathop {\max }\limits_{\{{\boldsymbol w}_k[n],\atop {\boldsymbol R}_s[n]\succeq \boldsymbol{0},{\boldsymbol q}[n]\}} & \frac{1}{N}\sum\limits_{n = 1}^N  {\sum\limits_{k = 1}^K \alpha_k{\log _2}(1+{\gamma _k}[n])}   \nonumber\\
\mathrm{s.t.}~&
{\boldsymbol{a}}^{\rm H}({\boldsymbol{{q}}}[n],{{\boldsymbol{m}}_j})\left(\sum\limits_{k = 1}^K {{{\boldsymbol{w}}_k}[n]{\boldsymbol{w}}_k^{\rm H}[n] + {{\boldsymbol{R}}_{ s}}} [n]\right){\boldsymbol{a}}({\boldsymbol{{q}}}[n],{{\boldsymbol{m}}_j}) \nonumber\\
&~~~~~~~~~~~~~~~~~~~~~~~~\ge d^2({\boldsymbol q}[n],{{\boldsymbol m}_j})\Gamma ,\forall j \in \mathcal J,n \in \mathcal N \label{P2-a beampattern}\\
~
&\sum\nolimits_{k = 1}^K {\|{{\boldsymbol{w}}_k}[n]\|{^2}}  + {\rm{tr}}({{\boldsymbol{R}}_{ s}}[n]) \le {P_{\max }}, \forall n\in \mathcal N  \label{P2-b tras power}\\
~
&(11),~(12),~\text{and}~(13),\nonumber
\end{align}
\end{subequations}
where ${\gamma _k}[n]$ denotes the received SINR by user $k$ at time slot $n$ given in \eqref{SINR}. Notice that problem (P2) is even more difficult to be solved than (P1) due to the involvement of more UAV trajectory variables. We will address problem (P2) in Section V. 

\subsection{Feasibility Checking for (P1) and (P2)}

Before proceeding to solve problems (P1) and (P2), we first check their feasibility. This is equivalent to solving the following two feasibility problems, respectively.
\begin{align}
\text{(FP1)}: ~& {\rm{find }}~\{{{\boldsymbol{w}}_k}\},{\boldsymbol{R}}_{s},{\boldsymbol{q}}   \nonumber\\
\mathrm{s.t.}~&(10{\rm a}),~(10{\rm b}),~ {\text{and}}~{\boldsymbol{R}}_{s}\succeq \boldsymbol{0}.\nonumber
\end{align}
\begin{align}
\text{(FP2)}: ~& {\rm{find }}~\{{\boldsymbol w}_k[n], {\boldsymbol R}_s[n], {\boldsymbol q}[n]\}   \nonumber\\
\mathrm{s.t.} ~& {\boldsymbol{R}}_{s}[n ]\succeq \boldsymbol{0}, \forall n \in \mathcal{N}\label{FP2-a}\\
~
&(11),~(12),~(13),~ (14{\rm a}),~\text{and}~(14{\rm b}). \nonumber
\end{align}

First, we consider problem (FP1). It can be shown that solving problem (FP1) is equivalent to solving the following problem with only sensing signals, by setting ${\boldsymbol w}_k=0, \forall k \in \mathcal{K}$, in problem (FP1).
\begin{subequations}
\begin{align}
\text{(FP3)}: ~& {\rm{find }}~{{\boldsymbol{R}}_{ s}},{\boldsymbol{q}}   \nonumber\\
\mathrm{s.t.} ~&{\boldsymbol{a}}^{\rm H}({\boldsymbol{q}},{{\boldsymbol{m}}_j}) {{\boldsymbol{R}}_{ s}} {\boldsymbol{a}}({\boldsymbol{q}},{{\boldsymbol{m}}_j}) \ge {{d^{2 }({\boldsymbol q},{{\boldsymbol m}_j})}}\Gamma ,\forall j \in \mathcal{J} \label{FP3-a}\\
~
& {\rm{tr}}({{\boldsymbol{R}}_{s}}) \le {P_{\max }} \label{FP3-b}\\
~
&{\boldsymbol{R}}_{s}\succeq \boldsymbol{0}. \label{FP3-c}
\end{align}
\end{subequations}
The equivalence between (FP1) and (FP3) can be validated by noticing the fact that for any feasible solution $\{{\boldsymbol{w}}_k\}$, ${\boldsymbol R}_{ s}$, and ${\boldsymbol q}$ to (FP1), we can always construct an equivalent solution $  \{ {\tilde{{{\boldsymbol{w}}}}_k} = 0,\forall k\} $, ${\tilde {\boldsymbol{R}}}_{ s}= \sum\nolimits_{k = 1}^K {\tilde {{\boldsymbol{w}}}_k{\tilde {\boldsymbol{w}}}_k^{{\rm H}} + {\tilde{\boldsymbol{R}}}_{ s}} $, ${\tilde {\boldsymbol q}}= {\boldsymbol q}$ that is feasible for both (FP1) and (FP3). Next, problem (FP3) can be solved by first solving the following problem (FP4) via optimizing ${\boldsymbol R}_s$ under any given ${\boldsymbol q}$, and then adopting a 2D location search on $\boldsymbol{q}$ over the whole interested area. As long as there exists one location ${\boldsymbol q}$ such that problem (FP4) is feasible, it follows that problem (FP3) and thus problems (FP1) and (P1) are feasible. 
\begin{align}
\text{(FP4)}: ~& {\rm{find }}~{{\boldsymbol{R}}_{ s}}   \nonumber\\
\mathrm{s.t.} ~&(16{\rm a}),~(16{\rm b}),~\text{and}~(16{\rm c}).\nonumber
\end{align}
As problem (FP4) is a standard semi-definite program (SDP), it can be solved via standard convex optimization tools such as CVX \cite{MG2014}. Therefore, problem (FP1) is solved. 

Next, we consider problem (FP2) for the mobile UAV scenario. Notice that by solving (FP3), we already find a set of UAV locations ${{\mathcal{Q}}^{\rm h}}$ in the interested area such that the sensing constraints in \eqref{P1-a} can be ensured (i.e., problem (FP4) is feasible when the UAV is located at any location in set ${{\mathcal{Q}}^{\rm h}}$). Based on this, checking the feasibility of (FP2) is equivalent to checking whether there exists a feasible UAV trajectory in ${{\mathcal{Q}}^{\rm h}}$ connecting the initial and final locations while satisfying the flight speed constraints. This is implemented based on the graph theory \cite{JA1976}, as detailed in the following. 

More specifically, solving (FP2) is equivalent to checking  whether there exists a path between the two nodes $\hat{\boldsymbol q}^{\rm I}$ and $\hat{\boldsymbol q}^{\rm F}$ with distance less than ${V_{\max }}N$, which corresponds to a typical reachability problem in graph theory that can be solved via the depth-first search (DFS) method \cite{JA1976}. Towards this end, we first construct an undirected graph with respect to the nodes in ${{\mathcal{Q}}^{\rm h}}$. Specifically, for each node in ${{\mathcal{Q}}^{\rm h}}$, we calculate its distance with all other nodes. If the distance between any two nodes is smaller than ${V_{\max }}$, then we connect them with an edge weighted by their distance. Next, we employ the DFS method on the initial location $\hat{\boldsymbol q}^{\rm I}$ to get a set ${{\mathcal{Q}}^{\rm c}}$ of connected components starting from $\hat{\boldsymbol q}^{\rm I}$. Note that the details of the DFS method can be referred to \cite{JA1976}, which is omitted for brevity. In the constructed set ${{\mathcal{Q}}^{\rm c}}$, any two nodes are reachable (i.e., at least one path exists between them). Based on the constructed ${{\mathcal{Q}}^{\rm c}}$, it can be concluded that if the final location $\hat{\boldsymbol q}^{\rm F}$ is included in ${{\mathcal{Q}}^{\rm c}}$, then there exists at least one feasible trajectory solution to problem (FP2) and thus (P2). Otherwise, problems  (FP2) and (P2) are infeasible. Therefore, problem (FP2) is solved.

In the following sections, we focus on the case when problems (P1) and (P2) are feasible.

\section{Proposed Solution to Problem (P1) for Quasi-stationary UAV Scenario}
In this section, we propose to find a high-quality solution to problem (P1), in which we adopt the SCA to optimize the information and sensing beamforming vectors (i.e., $\{\boldsymbol{w}_k\}$ and $\boldsymbol{R}_s$) under any given UAV deployment location $\boldsymbol q$, and then use the 2D search to find an optimized $\boldsymbol q$ that achieves the maximum weighted sum rate value. In the following, we focus our study on the optimization of $\{{\boldsymbol{w}}_k\}$ and ${{\boldsymbol{R}}_s}$ under given $\boldsymbol q$, which corresponds to solving the following problem:
\begin{align}
\text{(P3)}:\mathop {\max }\limits_{\{{{\boldsymbol{w}}_k}\},{{\boldsymbol{R}}_s}\succeq \boldsymbol{0}} ~& {\sum\nolimits_{k = 1}^K} \alpha_k{\log _2}\left(1 + \frac{{|{{\boldsymbol{h}}_k^{\rm{H}}({\boldsymbol{q}},{{\boldsymbol{u}}_k}){{\boldsymbol{w}}_k}}|^2}}{{\sum\nolimits_{i=1, i \ne k}^K {|{{\boldsymbol{h}}_k^{\rm{H}}({\boldsymbol{q}},{{\boldsymbol{u}}_k})}{{\boldsymbol{w}}_i}|^2}  + {{\boldsymbol{h}}_k^{\rm{H}}({\boldsymbol{q}},{{\boldsymbol{u}}_k})}{{\boldsymbol{R}}_s}{{\boldsymbol{h}}_k({\boldsymbol{q}},{{\boldsymbol{u}}_k})} + {\sigma ^2}}}\right)   \nonumber\\
\mathrm{s.t.}~&~(10{\rm a})~ {\text{and}}~(10{\rm b}).\nonumber
\end{align}
Note that the objective in (P3) is still non-convex, thus making it difficult to be optimally solved in general. In the following, we deal with this issue by using the techniques of SDR and SCA. 

First, we define $ {{\boldsymbol{W}}_k} = {{\boldsymbol{w}}_k}{\boldsymbol{w}}_k^{\rm H}$, where ${\boldsymbol{W}}_k\succeq 0$ and ${\rm{rank}}({{\boldsymbol{W}}_k}) \le 1$. By replacing ${{\boldsymbol{w}}_k}{\boldsymbol{w}}_k^{\rm H}$ as $ {{\boldsymbol{W}}_k}$, problem (P3) is reformulated as 
\begin{subequations}
\begin{align}
\text{(P4)}:\mathop {\max }\limits_{\{ {{\boldsymbol{W}}_k} \succeq \boldsymbol{0}\},{{\boldsymbol{R}}_s}\succeq \boldsymbol{0}} ~&  {\sum\nolimits_{k = 1}^K}\alpha_k {\hat r}_k(\{{{\boldsymbol{W}}_k}\},{{\boldsymbol{R}}_s})   \nonumber\\
\mathrm{s.t.}~&
{\boldsymbol{a}}^{\rm H}({\boldsymbol{q}},{{\boldsymbol{m}}_j})\left(\sum\nolimits_{k = 1}^K {{\boldsymbol{W}}_k + {{\boldsymbol{R}}_s}} \right){\boldsymbol{a}}({\boldsymbol{q}},{{\boldsymbol{m}}_j}) \ge {{d^{2 }({\boldsymbol q},{{\boldsymbol m}_j})}}\Gamma  ,\forall j \in \mathcal J \label{P4 beampattern}\\
~
&\sum\nolimits_{k = 1}^K {{\rm{tr}}({{\boldsymbol{W}}_k})}  + {\rm{tr}}({{\boldsymbol{R}}_s}) \le {P_{\max }}  \label{P4 power}\\
%~
%&{\boldsymbol{W}}_k \succeq 0,\forall k \in \mathcal{K}\tag{17c}\\
~
&{\rm{rank}}({{\boldsymbol{W}}_k}) \le 1 ,\forall k \in \mathcal{K},\label{P4 rank}
\end{align}
\end{subequations}
where ${\hat r}_k(\{{{\boldsymbol{W}}_k}\},{{\boldsymbol{R}}_s})={\log _2}\left(1 + \frac{{{\rm{tr}}\left({{\boldsymbol{h}}_k({\boldsymbol{q}},{{\boldsymbol{u}}_k})}{{\boldsymbol{h}}_k^{\rm{H}}({\boldsymbol{q}},{{\boldsymbol{u}}_k})}{{\boldsymbol{W}}_k}\right)}}{{\sum\nolimits_{i=1,i \ne k}^K {{\rm tr}\left({{\boldsymbol{h}}_k({\boldsymbol{q}},{{\boldsymbol{u}}_k})}{{\boldsymbol{h}}_k^{\rm{H}}({\boldsymbol{q}},{{\boldsymbol{u}}_k})}{{\boldsymbol{W}}_i}\right)}  + {\rm tr}\left({{\boldsymbol{h}}_k({\boldsymbol{q}},{{\boldsymbol{u}}_k})}{{\boldsymbol{h}}_k^{\rm{H}}({\boldsymbol{q}},{{\boldsymbol{u}}_k})}{{\boldsymbol{R}}_s}\right) + {\sigma ^2}}}\right) $. Note that problem (P4) is still non-convex, due to the non-concave objective function and the rank constraints in \eqref{P4 rank}.

Next, we use the SCA to address problem (P4) by approximating the non-concave objective function as a concave one, which is implemented in an iterative manner. Consider each iteration $l \ge 1$, in which the local point is denoted by $\{{\boldsymbol W}_k^{(l)}\}$ and ${\boldsymbol R}_s^{(l)}$. It then follows that
\begin{align}
&{\hat r}_k(\{{{\boldsymbol{W}}_k}\},{{\boldsymbol{R}}_s}) \nonumber\\
&={\log _2}\left(\sum\nolimits_{i = 1}^K {{\rm tr}\left({{\boldsymbol{h}}_k({\boldsymbol{q}},{{\boldsymbol{u}}_k})}{{\boldsymbol{h}}_k^{\rm{H}}({\boldsymbol{q}},{{\boldsymbol{u}}_k})}{{\boldsymbol{W}}_i}\right)}  + {\rm tr}\left({{\boldsymbol{h}}_k({\boldsymbol{q}},{{\boldsymbol{u}}_k})}{{\boldsymbol{h}}_k^{\rm{H}}({\boldsymbol{q}},{{\boldsymbol{u}}_k})}{{\boldsymbol{R}}_s}\right) + {\sigma ^2}\right) \nonumber\\
& ~~~~- {\log _2}\left(\sum\nolimits_{i=1,i \ne k}^K {{\rm{tr}}\left({{\boldsymbol{h}}_k({\boldsymbol{q}},{{\boldsymbol{u}}_k})}{{\boldsymbol{h}}_k^{\rm{H}}({\boldsymbol{q}},{{\boldsymbol{u}}_k})}{{\boldsymbol{W}}_i}\right)}  + {\rm{tr}}\left({{\boldsymbol{h}}_k({\boldsymbol{q}},{{\boldsymbol{u}}_k})}{{\boldsymbol{h}}_k^{\rm{H}}({\boldsymbol{q}},{{\boldsymbol{u}}_k})}{{\boldsymbol{R}}_s}\right) + {\sigma ^2}\right) \label{original rate express} \\
& \ge {\log _2}\left(\sum\nolimits_{i = 1}^K {{\rm tr}\left({{\boldsymbol{h}}_k({\boldsymbol{q}},{{\boldsymbol{u}}_k})}{{\boldsymbol{h}}_k^{\rm{H}}({\boldsymbol{q}},{{\boldsymbol{u}}_k})}{{\boldsymbol{W}}_i}\right)}  + {\rm tr}\left({{\boldsymbol{h}}_k({\boldsymbol{q}},{{\boldsymbol{u}}_k})}{{\boldsymbol{h}}_k^{\rm{H}}({\boldsymbol{q}},{{\boldsymbol{u}}_k})}{{\boldsymbol{R}}_s}\right) + {\sigma ^2}\right) \nonumber\\
&~~~~-\left( a_k^{(l)}+ \sum\nolimits_{i=1,i \ne k}^K {\rm tr}\big({{\boldsymbol{B}}_k^{(l)}({{\boldsymbol{W}}_i} \!-\! {\boldsymbol{W}}_i^{(l)})}\big)  + {\rm tr}\big({\boldsymbol{B}}_k^{(l)}({{\boldsymbol{R}}_s}\! -\! {\boldsymbol{R}}_s^{(l)})\big) \right) \triangleq \bar r_k^{(l)}(\{{{\boldsymbol{W}}_k}\}, {\boldsymbol{R}}_s),\label{lb rate}
\end{align}
where 
\begin{align}\label{lb a}
a_k^{(l)}\! =\! {\log _2}\left(\sum\nolimits_{i=1,i \ne k}^K {{\rm{tr}}\big({{\boldsymbol{h}}_k({\boldsymbol{q}},{{\boldsymbol{u}}_k})}{{\boldsymbol{h}}_k^{\rm{H}}({\boldsymbol{q}},{{\boldsymbol{u}}_k})}{\boldsymbol{W}}_i^{(l)}\big)} \! +\! {\rm{tr}}\left({{\boldsymbol{h}}_k({\boldsymbol{q}},{{\boldsymbol{u}}_k})}{{\boldsymbol{h}}_k^{\rm{H}}({\boldsymbol{q}},{{\boldsymbol{u}}_k})}{\boldsymbol{R}}_s^{(l)}\right)\! +\! {\sigma ^2}\right),
\end{align}
\begin{align}\label{lb b}
{\boldsymbol{B}}_k^{(l)} = \frac{{{{\log }_2}(e){{\boldsymbol{h}}_k({\boldsymbol{q}},{{\boldsymbol{u}}_k})}{{\boldsymbol{h}}_k^{\rm{H}}({\boldsymbol{q}},{{\boldsymbol{u}}_k})}}}{{\sum\nolimits_{i=1,i \ne k}^K {{\rm{tr}}\left({{\boldsymbol{h}}_k({\boldsymbol{q}},{{\boldsymbol{u}}_k})}{{\boldsymbol{h}}_k^{\rm{H}}({\boldsymbol{q}},{{\boldsymbol{u}}_k})}{\boldsymbol{W}}_i^{(l)}\right)}  + {\rm{tr}}\left({{\boldsymbol{h}}_k({\boldsymbol{q}},{{\boldsymbol{u}}_k})}{{\boldsymbol{h}}_k^{\rm{H}}({\boldsymbol{q}},{{\boldsymbol{u}}_k})}{\boldsymbol{R}}_s^{(l)}\right) + {\sigma ^2}}}.
\end{align}
Here, the formula in \eqref{original rate express} has a concave-minus-concave form, and \eqref{lb rate} follows by implementing the first-order Taylor expansion on the second term in \eqref{original rate express}. Accordingly, by replacing ${\hat r}_k(\{{{\boldsymbol{W}}_k}\},{{\boldsymbol{R}}_s})$ as its lower bound  $\bar r_k^{(l)}(\{{{\boldsymbol{W}}_k}\}, {\boldsymbol{R}}_s)$, problem (P4) is approximated as the following problem (P5.$l$) in the $l$-th iteration of SCA. 
\begin{align}
\text{(P5.$l$)}:\mathop {\max }\limits_{\{ {{\boldsymbol{W}}_k} \succeq \boldsymbol{0}\},{{\boldsymbol{R}}_s}\succeq \boldsymbol{0}} ~&  {\sum\nolimits_{k = 1}^K}\alpha_k \bar r_k^{(l)}(\{{{\boldsymbol{W}}_k}\},{\boldsymbol{R}}_s)     \nonumber\\
\mathrm{s.t.}~& (17{\rm a}), ~(17{\rm b}),~\text{and}~(17{\rm c}). \nonumber
\end{align}

Next, we deal with the non-convex rank constraints in \eqref{P4 rank} in problem (P5.$l$) via the idea of SDR. In particular, we relax the rank constraints in \eqref{P4 rank} and express the relaxed problem as (SDR5.$l$). Note that problem (SDR5.$l$) is a convex SDP and thus can be solved optimally by convex optimization solvers such as CVX. Let $ \{ {{\boldsymbol{\dot W}}^{(l)}_k}\}$ and ${{\boldsymbol{\dot R}}^{(l)}_s}$ denote the obtained solution to problem (SDR5.$l$). In particular, if ${\rm rank}({{\boldsymbol{\dot W}}^{(l)}_k}) \le 1, \forall k \in \mathcal{K}$, then the SDR is tight, i.e., $ \{ {{\boldsymbol{\dot W}}^{(l)}_k}\}$ and ${{\boldsymbol{\dot R}}^{(l)}_s}$ are also optimal for problem (P5.$l$). Otherwise, we need to further construct the rank-one solution of $\{{\boldsymbol W}_k\}$ for (P5.$l$) via additional procedures such as the Gaussian randomization \cite{ZL2014}. Fortunately, the following proposition shows that there always exists an optimal rank-one solution of $\{{\boldsymbol W}_k\}$ to (SDR5.$l$), and thus the Gaussian randomization is not needed to solve problem (P5.$l$).

\begin{proposition}\label{rankone}
There always exists a globally optimal solution to problem (SDR5.$l$), denoted as $ \{ {{\boldsymbol{\bar W}}^{(l)}_k}\}$ and ${{\boldsymbol{\bar R}}^{(l)}_s} $, such that
\begin{align}
{\rm{rank}}({{\boldsymbol{\bar W}}^{(l)}_k}) = 1, \forall k \in \mathcal K . \nonumber
\end{align}

Proof: See Appendix A.
\end{proposition}

Therefore, by iteratively solving problem (SDR5.$l$) and thus (P5.$l$), we can obtain a series of solutions $ \{ {{\boldsymbol{\bar W}}^{(l)}_k}\}$'s and ${{\boldsymbol{\bar R}}^{(l)}_s}$'s, which lead to monotonically non-decreasing objective values for (P4) and thus (P3). Therefore, the convergence of the SCA-and-SDR-based algorithm for solving problem (P3) is ensured.

%The algorithm to solve (P3) is summarized in Algorithm 1, which is ensured to converge. 

%\begin{algorithm}[h]
%\caption{Overall Algorithm for (P3)}
%\label{alg:Framwork}
%\begin{algorithmic}[1] %这个1 表示每一行都显示数字
%\STATE Initialize the information beamforming vectors $\{{\boldsymbol w}^{(0)}_k\}$\\
%and the dedicated sensing covariance matrix ${\boldsymbol R}^{(0)}_s $. Let $r=1$.
%\REPEAT
%\STATE Under given $\boldsymbol q$, solve (SDR5.r) at local point $\{{\boldsymbol W}^{(r-1)}_k\}$, ${\boldsymbol R}^{(r-1)}_s $ to obtain $ \{ {{\boldsymbol{\dot W}}^{(r)}_k}\}$ and ${{\boldsymbol{\dot R}}^{(r)}_s}$ .
%\STATE Reconstruct rank-one solutions $ \{ {{\boldsymbol{\bar W}}^{(r)}_k}\}$ and ${{\boldsymbol{\bar R}}^{(r)}_s} $ via proposition 3.1.
%\STATE Update $\{{\boldsymbol W}^{(r)}_k\}=\{ {{\boldsymbol{\bar W}}^{(r)}_k}\}$, ${\boldsymbol R}^{(r)}_s={{\boldsymbol{\bar R}}^{(r)}_s}$,
  %  $r=r+1$.
%\UNTIL The increase of the objective value is below a threshold $ \varsigma$. 
%\end{algorithmic}
%\end{algorithm}

\section{Proposed Solution to Problem (P2) for Mobile UAV Scenario}
This section addresses problem (P2) for the mobile UAV scenario by using the alternating optimization together with SCA. Specifically, the alternating optimization based approach is implemented in an iterative manner. In each iteration, we first optimize the information beamforming vectors $\{{\boldsymbol w}_k[n]\} $ and the sensing covariance matrix $\{{\boldsymbol R}_s[n]\} $ with given UAV trajectory $\{{\boldsymbol q}[n]\} $, and then optimize the UAV trajectory $\{{\boldsymbol q}[n]\} $ with updated  $\{{\boldsymbol w}_k[n]\} $ and $\{{\boldsymbol R}_s[n]\} $, as detailed in the sequel.

\subsection{Transmit Information and Sensing Beamforming Optimization}

First, we consider the optimization of information beamforming vectors $\{{\boldsymbol w}_k[n]\} $ and sensing covariance matrix $\{{\boldsymbol R}_s[n]\} $ in problem (P2) under given trajectory $\{{\boldsymbol q}[n]\} $, for which the optimization problem becomes
\begin{align}
\text{(P6)}:\mathop {\max }\limits_{\{{\boldsymbol w}_k[n],\atop {\boldsymbol R}_s[n]\succeq \boldsymbol{0}\}} & \sum\limits_{n = 1}^N {\sum\limits_{k = 1}^K} \alpha_k {\log _2} (1+{\gamma _k}[n]) \nonumber\\
\mathrm{s.t.}~&~(14{\rm a})~ {\text{and}}~(14{\rm b}).\nonumber
\end{align}
It is observed from problem (P6) that the optimization of $\{{\boldsymbol w}_k[n]\} $ and $\{{\boldsymbol R}_s[n]\}$ over different time slots $n$'s can be decoupled. In this case, problem (P6) can be equivalently decomposed into a number of $N$ subproblems each for one time slot $n$. Note that each subproblem at slot $n$ has the same form of problem (P3) with the UAV location given as ${\boldsymbol q}[n]$ (instead of $\boldsymbol {q}$ in (P3)). As a result, problem (P6) can be solved equivalently by implementing the SCA-and-SDR-based algorithm in Section IV $N$ times each for one slot, for which the details are omitted for brevity.

\subsection{UAV trajectory Optimization}

Next, we optimize the UAV's trajectory $\{{\boldsymbol q}[n]\}$ under given $\{{\boldsymbol w}_k[n]\}$ and $\{{\boldsymbol R}_s[n]\}$, for which the optimization problem is expressed as follows, in which the channel vectors ${\boldsymbol{h}}_k({\boldsymbol{{q}}}[n],{{\boldsymbol{u}}_k})$'s are explicitly rewritten as $\sqrt {{\beta }/({{{H^2} + \|{\boldsymbol{q}}[n] - {{\boldsymbol{u}}_k}\|{^2}}})} {\boldsymbol{a}}({\boldsymbol{q}}[n],{{\boldsymbol{u}}_k})$ based on \eqref{channel}. 
\begin{align}
\text{(P7)}:\mathop {\max }\limits_{\{{\boldsymbol q}[n]\}} ~& \sum\limits_{n = 1}^N {\sum\limits_{k = 1}^K} \alpha_k {\log _2}\left(1 + \frac{{\frac{{\beta |{\boldsymbol{a}}^{\rm H}({\boldsymbol{{q}}}[n],{{\boldsymbol{u}}_k}){{\boldsymbol{w}}_k}[n]{|^2}}}{{{H^2} + \|{\boldsymbol{q}}[n] - {{\boldsymbol{u}}_k}\|{^2}}}}}{{\sum\nolimits_{i=1,i \ne k}^K {\frac{{\beta |{\boldsymbol{a}}^{\rm H}({\boldsymbol{{q}}}[n],{{\boldsymbol{u}}_k}){{\boldsymbol{w}}_k}[n]{|^2}}}{{{H^2} + \|{\boldsymbol{q}}[n] - {{\boldsymbol{u}}_k}\|{^2}}}}  + \frac{{\beta {\boldsymbol{a}}^{\rm H}({\boldsymbol{{q}}}[n],{{\boldsymbol{u}}_k}){{\boldsymbol{R}}_s}[n]{\boldsymbol{a}}({\boldsymbol{{q}}}[n],{{\boldsymbol{u}}_k})}}{{{H^2} + \|{\boldsymbol{q}}[n] - {{\boldsymbol{u}}_k}\|{^2}}} + {\sigma ^2}}}\right)   \nonumber\\
\mathrm{s.t.}~&
 {\boldsymbol{a}}^{\rm H}({\boldsymbol{{q}}}[n],{{\boldsymbol{m}}_j})\left(\sum\limits_{k = 1}^K {{{\boldsymbol{w}}_k}[n]{\boldsymbol{w}}_k^{\rm H}[n] + {{\boldsymbol{R}}_s}} [n]\right){\boldsymbol{a}}({\boldsymbol{{q}}}[n],{{\boldsymbol{m}}_j}) \nonumber \\
&~~~~~~~~~~~~\ge \Gamma({{{{{H^2} + \|{\boldsymbol{q}}[n] - {\boldsymbol{m}}_j\|{^2}}}}}) ,\forall j \in \mathcal J,n \in \mathcal N \label{P7 beampattern}\\
~
&(11),~(12),~\text{and}~(13). \nonumber
\end{align}

Note that problem (P7) is difficult to be solved optimally due to the non-concave objective function and the non-convex constraints in \eqref{P7 beampattern}, in which the trajectory variables $\{{\boldsymbol q}[n]\}$ are involved in the steering vector ${\boldsymbol{a}}({\boldsymbol{q}}[n],{{\boldsymbol{u}}_k})$. To deal with the highly non-convex problem (P7), in the following, we propose a trust-region-based SCA algorithm. 

For notational convenience, we express ${{\boldsymbol{W}}_i}[n] = {{\boldsymbol{w}}_i}[n]{\boldsymbol{w}}_i^{\rm H}[n]$ and ${\boldsymbol{G}}[n] = \sum\nolimits_{k = 1}^K {{{\boldsymbol{w}}_k}[n]{\boldsymbol{w}}_k^H[n]} + {{\boldsymbol{R}}_s} [n]$. Accordingly, we denote the entries in the $p$-th row and $q$-th column of ${{\boldsymbol W}_i}[n]$, ${{\boldsymbol{R}}_s}[n]$, and ${\boldsymbol{G}}[n]$ as $\big[{{\boldsymbol W}_i}[n]\big]_{p,q}$, $\big[{{\boldsymbol{R}}_s}[n]\big]_{p,q}$, and $\big[{\boldsymbol{G}}[n]\big]_{p,q}$, whose magnitudes are denoted by $\big|\big[{{\boldsymbol W}_i}[n]\big]_{p,q}\big|$, $\big|\big[{{\boldsymbol{R}}_s}[n]\big]_{p,q}\big|$, and $\big|\big[{\boldsymbol{G}}[n]\big]_{p,q}\big|$, and phases denoted by $\theta _{p,q}^{{\rm W}_i}[n]$, $\theta _{p,q}^{\rm R}[n]$, and $\theta _{p,q}^{\rm G}[n]$, respectively. 
%(the same rules are applied to ${{\boldsymbol R}_s}[n]$ as well). 
Then we re-express the objective function of (P7) as
\begin{align}\label{P7 objective trans}
{\hat R_k}[n] &= {\log _2}\left(\sum\limits_{i = 1}^K \eta \left({{\boldsymbol{W}}_i}[n],{d}({\boldsymbol{q}}[n],{{\boldsymbol{u}}_k})\right) + \mu  \left({{\boldsymbol{R}}_s}[n],{d}({\boldsymbol{q}}[n],{{\boldsymbol{u}}_k})\right) + \frac{{{\sigma ^2}}}{\beta }d^2({\boldsymbol{q}}[n],{{\boldsymbol{u}}_k})\right) \nonumber\\
&-{\log _2}\left(\sum\limits_{i=1,i \ne k}^K \eta \left({{\boldsymbol{W}}_i}[n],{d}({\boldsymbol{q}}[n],{{\boldsymbol{u}}_k})\right)\! + \!\mu  \left({{\boldsymbol{R}}_s}[n],{d}({\boldsymbol{q}}[n],{{\boldsymbol{u}}_k})\right) \!+ \!\frac{{{\sigma ^2}}}{\beta }d^2({\boldsymbol{q}}[n],{{\boldsymbol{u}}_k})\right), 
\end{align}
where 
\begin{align}\label{eta}
&{\eta \left({{\boldsymbol{W}}_i}[n],{d}({\boldsymbol{q}}[n],{{\boldsymbol{u}}_k})\right)} \nonumber\\
&=\sum\limits_{\alpha {\rm{ = }}1}^M {\big[{{\boldsymbol W}_i}[n]\big]_{\alpha ,\alpha }}  + 2\sum\limits_{p = 1}^M {\sum\limits_{q = p + 1}^M {\big|\big[{{\boldsymbol W}_i}[n]\big]_{p,q}\big|{\rm{cos}}\left(\theta _{p,q}^{{\rm W}_i}[n]{\rm{ + }}2\pi \frac{d}{\lambda }(q - p)\frac{H}{{d({\boldsymbol{q}}[n],{{\boldsymbol{u}}_k})}}\right)} } ,
\end{align}
\begin{align}\label{mu}
&\mu \left({{\boldsymbol{R}}_s}[n],{d}({\boldsymbol{q}}[n],{{\boldsymbol{u}}_k})\right)\nonumber \\
&= \sum\limits_{\alpha {\rm{ = }}1}^M {\big[{{\boldsymbol{R}}_s}[n]\big]_{\alpha ,\alpha }}  + 2\sum\limits_{p = 1}^M {\sum\limits_{q = p + 1}^M {\big|\big[{{\boldsymbol{R}}_s}[n]\big]_{p,q}\big|{\rm{cos}}\left(\theta _{p,q}^{\rm R}[n]{\rm{ + }}2\pi \frac{d}{\lambda }(q - p)\frac{H}{{d^{}({\boldsymbol{q}}[n],{{\boldsymbol{u}}_k})}}\right)} }.
\end{align}
Note that the derivation of \eqref{P7 objective trans} is presented in Appendix B. Similarly, we also re-express the non-convex constraints in \eqref{P7 beampattern} as
\begin{align}\label{P7 beampattern trans}
\sum\limits_{\alpha {\rm{ = }}1}^M {{\big[{\boldsymbol{G}}[n]\big]_{\alpha ,\alpha }}}  + 2\sum\limits_{p = 1}^M {\sum\limits_{q = p + 1}^M {\big|\big[{\boldsymbol{G}}[n]\big]_{p ,q }\big|{\rm{cos}}\left(\theta _{p,q}^{\rm G}[n]{\rm{ + }} \frac{2\pi d(q - p)H}{\lambda {{d}({{\boldsymbol{q}}}[n],{{\boldsymbol{m}}_j})}}\right)} }  \ge \Gamma d^2({{\boldsymbol{q}}}[n],{{\boldsymbol{m}}_j}). 
\end{align}

Now, we are ready to present the trust-region-based SCA algorithm, which is implemented in an iterative manner. Consider a particular iteration $l \ge 1$ with local trajectory point $ {{\boldsymbol{q}}^{(l)}}[n]$. First, we deal with the non-concave objective function in \eqref{P7 objective trans}, which is approximated as follows based on its first-order Taylor expansion. 
\begin{align}\label{P7 obj linear}
{{\hat R}_k}[n] \approx {{\bar R}^{(l)}_k}[n]\buildrel \Delta \over = {c^{(l)}_k}[n] + {{\boldsymbol d}^{(l)}_k}^{\rm H}[n]({\boldsymbol{q}}[n] - {{\boldsymbol{q}}^{(l)}}[n]),
\end{align}
where 
\begin{align}\label{c}
{{c^{(l)}_k}}[n] &= {\log _2}\left(\sum\limits_{i = 1}^K \eta \big({{\boldsymbol{W}}_i}[n],{d}({{\boldsymbol{q}}^{(l)}}[n],{{\boldsymbol{u}}_k})\big) + \mu  \big({{\boldsymbol{R}}_s}[n],{d}({{\boldsymbol{q}}^{(l)}}[n],{{\boldsymbol{u}}_k})\big) + \frac{{{\sigma ^2}}}{\beta }d^2({{\boldsymbol{q}}^{(l)}}[n],{{\boldsymbol{u}}_k})\right) \nonumber\\
&\!\!\!\!\!\!-{\log _2}\!\left(\sum\limits_{i=1,i \ne k}^K \!\!\! \eta \big({{\boldsymbol{W}}_i}[n],{d}({{\boldsymbol{q}}^{(l)}}[n],{{\boldsymbol{u}}_k})\big) \!+\! \mu  \big({{\boldsymbol{R}}_s}[n],{d}({{\boldsymbol{q}}^{(l)}}[n],{{\boldsymbol{u}}_k})\big) \!+\! \frac{{{\sigma ^2}}}{\beta }d^2({{\boldsymbol{q}}^{(l)}}[n],{{\boldsymbol{u}}_k})\right)\!,
\end{align}

\begin{align}\label{d}
{{\boldsymbol{d}}^{(l)}_k}[n] &= \frac{{{{\log }_2}(e)}}{{{e_k}[n]}}\!\left(\sum\limits_{i = 1}^K \gamma \big({{\boldsymbol{W}}_i}[n],{d}({{\boldsymbol{q}}^{(l)}}[n],{{\boldsymbol{u}}_k})\big) \!+\! \omega  \big({{\boldsymbol{R}}_s}[n],{d}({{\boldsymbol{q}}^{(l)}}[n],{{\boldsymbol{u}}_k})\big) + \frac{{{\sigma ^2}}}{\beta }({{\boldsymbol{q}}^{(l)}}[n] - {{\boldsymbol{u}}_k})\!\! \right) \nonumber\\
 &\!\!\!\!\!\!\!\!- \frac{{{{\log }_2}(e)}}{{{f_k}[n]}}\!\! \left(\!\sum\limits_{i=1,i \ne k}^K \!\!\!\!\gamma \left({{\boldsymbol{W}}_i}[n],{d}({{\boldsymbol{q}}^{(l)}}[n],{{\boldsymbol{u}}_k})\right)\! +\! \omega  \big({{\boldsymbol{R}}_s}[n],{d}({{\boldsymbol{q}}^{(l)}}[n],{{\boldsymbol{u}}_k})\big) \!+\! \frac{{{\sigma ^2}}}{\beta }({{\boldsymbol{q}}^l}[n] - {{\boldsymbol{u}}_k})\right)\!\!,\!\!
\end{align}

with
\begin{align}\label{gamma}
\begin{array}{l}
{\gamma \left({{\boldsymbol{W}}_i}[n],{d}({{\boldsymbol{q}}^{(l)}}[n],{{\boldsymbol{u}}_k})\right)}\\
=\sum\limits_{p = 1}^M {\sum\limits_{q = p + 1}^M {4\pi \big|\big[{{\boldsymbol W}_i}[n]\big]_{p,q}\big|\sin \left(\theta _{p,q}^{{\rm W}_i}[n] + 2\pi \frac{d}{\lambda }(q - p)\frac{H}{{{d}({{\boldsymbol{q}}^{(l)}}[n],{{\boldsymbol{u}}_k})}}\right)} } \frac{{d(q - p)H}}{{\lambda d^3({{\boldsymbol{q}}^{(l)}}[n],{{\boldsymbol{u}}_k})}}({{\boldsymbol{q}}^{(l)}}[n] - {{\boldsymbol{u}}_k}),
\end{array}
\end{align}

\begin{align}\label{omega}
\begin{array}{l}
\omega \left({{\boldsymbol{R}}_s}[n],{d}({{\boldsymbol{q}}^{(l)}}[n],{{\boldsymbol{u}}_k})\right)\\
=\sum\limits_{p = 1}^M {\sum\limits_{q = p + 1}^M {4\pi \big|\big[{{\boldsymbol{R}}_s}[n]\big]_{p,q}\big|\sin \left(\theta _{p,q}^{\rm R}[n] + 2\pi \frac{d}{\lambda }(q - p)\frac{H}{{{d}({{\boldsymbol{q}}^{(l)}}[n],{{\boldsymbol{u}}_k})}}\right)} } \frac{{d(q - p)H}}{{\lambda d^3({{\boldsymbol{q}}^{(l)}}[n],{{\boldsymbol{u}}_k})}}({{\boldsymbol{q}}^{(l)}}[n] - {{\boldsymbol{u}}_k}), 
\end{array}
\end{align}

\begin{align}\label{e}
e_k[n] = \sum\limits_{i = 1}^K \eta \left({{\boldsymbol{W}}_i}[n],{d}({{\boldsymbol{q}}^{(l)}}[n],{{\boldsymbol{u}}_k})\right) + \mu  \left({{\boldsymbol{R}}_s}[n],{d}({{\boldsymbol{q}}^{(l)}}[n],{{\boldsymbol{u}}_k})\right) + \frac{{{\sigma ^2}}}{\beta }d^2({{\boldsymbol{q}}^{(l)}}[n],{{\boldsymbol{u}}_k}),
\end{align}

\begin{align}\label{f}
f_k[n] =  \sum\limits_{i=1,i \ne k}^K \eta \left({{\boldsymbol{W}}_i}[n],{d}({{\boldsymbol{q}}^{(l)}}[n],{{\boldsymbol{u}}_k})\right) + \mu  \left({{\boldsymbol{R}}_s}[n],{d}({{\boldsymbol{q}}^{(l)}}[n],{{\boldsymbol{u}}_k})\right) + \frac{{{\sigma ^2}}}{\beta }d^2({{\boldsymbol{q}}^{(l)}}[n],{{\boldsymbol{u}}_k}).
\end{align}
%Here, the objective function in (23) is approximated to a linear form as in (26) for each iteration $l$. 
Next, we deal with the non-convex constraints in \eqref{P7 beampattern trans}. Similarly as for \eqref{P7 obj linear}, we approximate the left-hand-side of \eqref{P7 beampattern trans} based on its first-order Taylor expansion at local point $ {{\boldsymbol{q}}^{(l)}}[n]$, and accordingly re-express \eqref{P7 beampattern trans} as
\begin{align}\label{P7 beampattern convex}
{h^{(l)}_j}[n] + {{\boldsymbol i}^{(l)}_j}^{\rm H}[n]({\boldsymbol{q}}[n] - {\boldsymbol{ q}}^{(l)}[n]) \ge \Gamma ({H^2} + \|{\boldsymbol{q}}[n] - {\boldsymbol{m}}_j\|{^2}),
\end{align}
where
\begin{align}\label{h}
{h^{(l)}_j}[n] = \sum\limits_{\alpha {\rm{ = }}1}^M {\big[{\boldsymbol{G}}[n]\big]_{\alpha,\alpha }}  + 2\sum\limits_{p = 1}^M {\sum\limits_{q = p + 1}^M {\big|\big[{\boldsymbol{G}}[n]\big]_{p ,q }\big|{\rm{cos}}\left(\theta _{p,q}^{\rm G}[n]{\rm{ + }}2\pi \frac{d}{\lambda }(q - p)\frac{H}{{{d}({{\boldsymbol{q}}^{(l)}}[n],{{\boldsymbol{m}}_j})}}\right)} } , 
\end{align}
\begin{align}\label{i}
{{\boldsymbol{i}}^{(l)}_j}[n] \!=\! \sum\limits_{p = 1}^M\!\! {\sum\limits_{q = p + 1}^M\! {4\pi \big|\big[{\boldsymbol{G}}[n]\big]_{p ,q }\big|{\rm{sin}}\left(\theta _{p,q}^{\rm G}[n]{\rm{ + }}\frac{{2\pi d(q - p)H}}{{\lambda {d}({{\boldsymbol{q}}^{(l)}}[n],{{\boldsymbol{m}}_j})}}\right)} } \frac{{d(q - p)H}}{{\lambda d^3({{\boldsymbol{q}}^{(l)}}[n],{{\boldsymbol{m}}_j})}}({{\boldsymbol{q}}^{(l)}}[n] \!- \!{\boldsymbol{ m}}_j).
\end{align}

So far, we approximate the non-concave objective function in \eqref{P7 objective trans} as a linear one in \eqref{P7 obj linear}, and the non-convex constraints in \eqref{P7 beampattern trans} as convex ones in \eqref{P7 beampattern convex}. To ensure the approximation accuracy, we impose a series of trust region constraints as
\begin{align}\label{trust region}
\|{{\boldsymbol{q}}^{(l)}}[n] - {{\boldsymbol{q}}^{(l-1)}}[n]\| \le {\psi ^{(l)}}, \forall n\in \mathcal N,
\end{align}
where $\psi ^{(l)}$ denotes the radius of the trust region. 
\begin{algorithm}[h]
\caption{Overall Algorithm for solving problem (P2)}
\label{alg:Framwork}
\begin{algorithmic}[1] %这个1 表示每一行都显示数字
\STATE Initialize the information beamforming vectors $\{{\boldsymbol w}^{(0)}_k[n]\}$,\\
the dedicated sensing signal covariance matrix $\{{\boldsymbol R}^{(0)}_s[n]\} $,\\
and the UAV trajectory $\{{\hat {\boldsymbol q}}^{(0)}[n]\}$; set $o=1$.\\
\REPEAT
\STATE ****{\it Optimizing information and sensing beamforming}****
\STATE Solve problem (P6) under local point $\{{\boldsymbol w}^{(o-1)}_k[n],{\boldsymbol R}^{(o-1)}_s[n],{\hat{\boldsymbol q}}^{(o-1)}[n]\}$ to obtain $ \{ {{\boldsymbol{\dot W}}^{(o)}_k}[n],{{\boldsymbol{\dot R}}^{(o)}_s}[n]\}$.\\
\STATE Reconstruct $ \{ {{\boldsymbol{\bar W}}^{(o)}_k}[n],{{\boldsymbol{\bar R}}^{(o)}_s}[n]\} $ (and correspondingly $ \{ {{\boldsymbol{\bar w}}^{(o)}_k}[n]\}$) based on proposition 4.1.\\
\STATE ****{\it Optimizing the UAV trajectory}****
\STATE Let $l=1$, $\{{\boldsymbol q}^{(l-1)}[n]\}=\{{\hat{\boldsymbol q}}^{(o-1)}[n]\}$.
  \REPEAT
  \STATE Obtain $\{{\boldsymbol q}^{(l)*}[n]\}$ by solving problem (P8.$l$) under local point $\{{\boldsymbol q}^{(l-1)}[n],{{\boldsymbol{\bar W}}^{(o)}_k}[n],{{\boldsymbol{\bar R}}^{(o)}_s}[n]\}$.
  \IF{the objective value of probelm (P7) increases}
  \STATE $\{{\boldsymbol q}^{(l)}[n]\}=\{{\boldsymbol q}^{(l)*}[n]\} $,    $l=l+1$.
  \ELSE
  \STATE Execute $\psi^{(l)}=\psi^{(l)}/2$.
  \ENDIF
  \UNTIL $\psi^{(l)} < \hat \varsigma$
\STATE Update $\{{\hat{\boldsymbol q}}^{(o)}[n]\}=\{{\boldsymbol q}^{(l)}[n]\} $,
   $o=o+1$.
\UNTIL the increase of the objective value is below a threshold $\bar \varsigma$. 
\end{algorithmic}
\end{algorithm}

Finally, by replacing the non-concave objective function in \eqref{P7 objective trans} and the non-convex constraints \eqref{P7 beampattern trans} as their approximate forms in \eqref{P7 obj linear} and \eqref{P7 beampattern convex}, respectively, and adding the trust region constraints in \eqref{trust region}, we obtain the approximated convex version of problem (P7) in the $l$-th iteration as problem (P8.$l$) in the following, which can be optimally solved via CVX efficiently. 
\begin{align}
\text{(P8.$l$)}:\mathop {\max }\limits_{\{{\boldsymbol q}[n]\}} ~&  \sum\nolimits_{n = 1}^N {\sum\nolimits_{k = 1}^K \alpha_k {{{\bar R}^{(l)}_k}[n]} }    \nonumber\\
\mathrm{s.t.}~&
(34),~(37), ~(11),~(12),~\text{and}~(13). \nonumber
\end{align}

In summary, by solving a series of problems in (P8.$l$) over iteration $l$'s, we can obtain an optimized solution to problem (P7). Notice that theoretically, if the radius of trust region $\psi^{(l)}$ is chosen to be sufficiently small, then the convergence of the iteration can always be ensured \cite{AR2000}. In practical implementation, if the objective value of (P7) after solving (P8.$l$) in each iteration $l$ is not reduced as compared to that in the previous round, then we reduce the radius of the trust region as $\psi^{(l)}=\psi^{(l)}/2$ and resolve (P8.$l$) again. The iteration will terminate when $\psi^{(l)}$ is lower than a given threshold $\hat \varsigma $. The convergence of the trust-region-based SCA algorithm can be guaranteed, as shown in Fig. \ref{converge} in Section VI next. Finally, the overall algorithm for problem (P2) is presented in Algorithm 1.

\section{Numerical Results}
This section presents numerical results to validate the performance of our proposed UAV-enabled ISAC designs. In the simulation, we consider an area of $1~{\rm km} \times 1~{\rm km}$ with $K = 8$ ground users and $J=18$ sample locations in the interested sensing area, as shown in Fig. \ref{simulation setup}. Unless otherwise stated, we set the antenna spacing as $d=\lambda/2$, the number of antennas at the UAV as $M = 12$, and the beampattern gain threshold $\Gamma= 5{\rm e}^{-5}~(-43~{\rm dBm})$. We also set the UAV's maximum horizontal flight speed as ${\tilde V}_{\max } = 30~{\rm{m/s}}$, flight altitude as ${H} = 100~\rm{m}$, and maximum transmit power as ${P_{\max }} = 0.5 ~ \rm{W}$. In addition, we set the noise power at each user receiver as $\sigma _k^2 = -110~\rm{dBm}$, and the channel power gain at the reference distance ${d_0} = 1~ {\rm{m}}$ as ${{\beta}} = -60~ {\rm{dB}}$. Furthermore, we set the users' rate weights are $\alpha_k = 1,\forall k\in\mathcal K$, such that their sum rate is considered as the communication performance metric.

\subsection{Quasi-stationary UAV Scenario}
First, we consider the quasi-stationary UAV scenario, for which the following benchmark schemes are considered for performance comparison. 
\begin{itemize}
\item \textbf{Communication only}: The UAV designs its deployment location and transmit information and sensing beamforming to optimize the communication performance only, by ignoring the sensing requirements. This corresponds to solving problem (P1) by setting the sensing beampattern threshold as $\Gamma = 0$. 

\item \textbf{Sensing only}: The UAV only performs the sensing task without the communication needs. In this case, the UAV aims to maximize the minimum beampattern gain weighted by the squared distance $d^2({\boldsymbol q},{{\boldsymbol m}_j})$, by optimizing the deployment location $\boldsymbol q$ and the sensing covariance matrix ${{\boldsymbol{{R}}}_s}$. This corresponds to solving the following problem:
\begin{align}
\text{(P9)}:\mathop {\max }\limits_{ {{\boldsymbol R}_s}\succeq 0, {\boldsymbol q}} ~& \min_{j\in\mathcal{J}} ~\frac{1}{d^2({\boldsymbol q},{{\boldsymbol m}_j})}{\boldsymbol{a}}^{\rm H}({\boldsymbol{{q}}},{{\boldsymbol{m}}_j}) {{\boldsymbol{R}}_s}{\boldsymbol{a}}({\boldsymbol{{q}}},{{\boldsymbol{m}}_j}) \nonumber\\
\mathrm{s.t.} ~& {\rm{tr}}({{\boldsymbol{R}}_s}) \le {P_{\max }} ,\label{P9 power}
\end{align}
which can be solved similarly as problem (P1), for which the details are omitted for brevity.
\end{itemize}

\begin{figure}[h]
\centering
 \epsfxsize=1\linewidth
    \includegraphics[width=9cm]{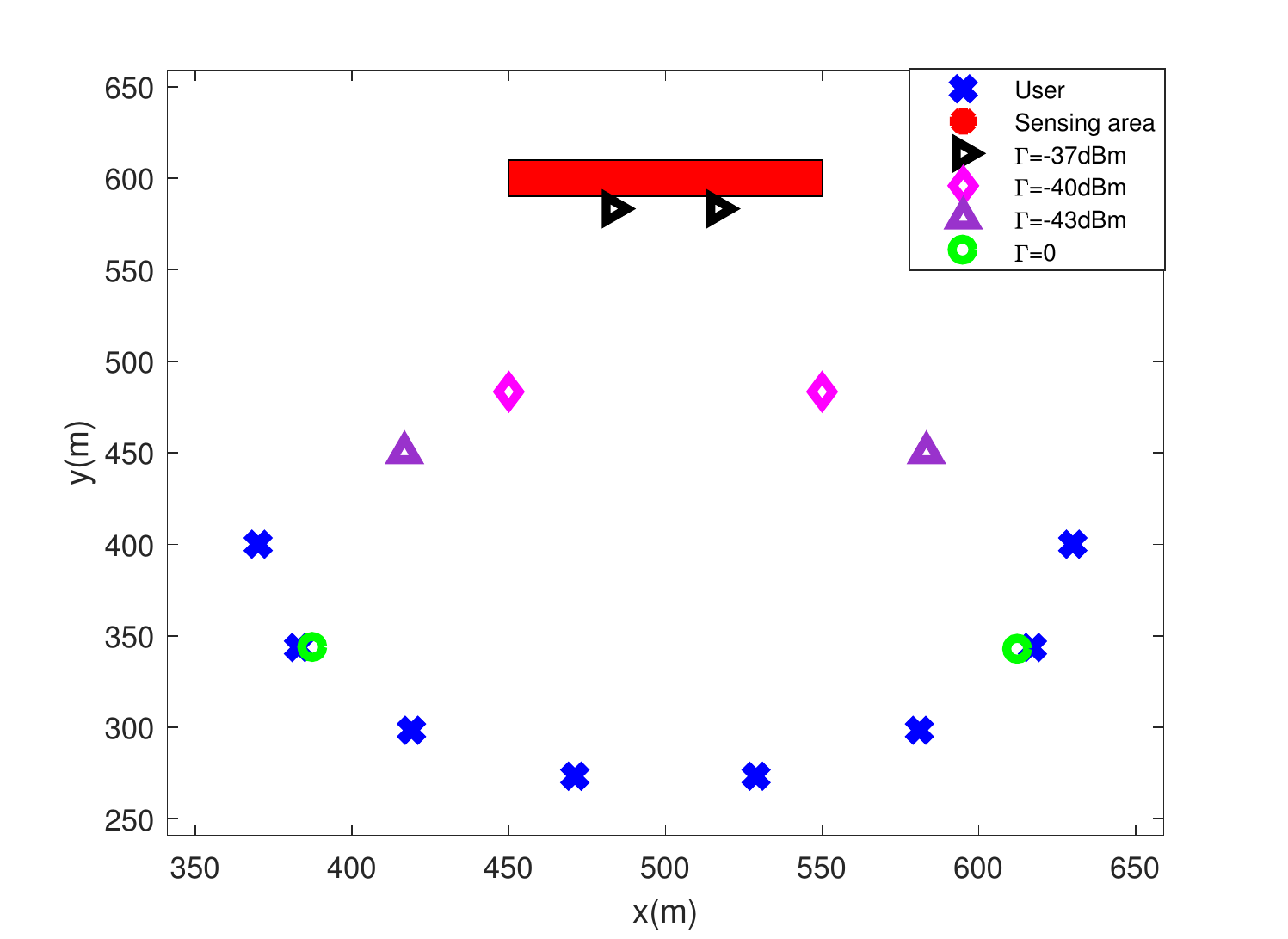}
\vspace{-4pt}
\caption{Simulation setup and the obtained deployment locations of the UAV under the different values of $\Gamma$.}
\label{simulation setup}
\end{figure}

\begin{figure}[h]
\centering
 \epsfxsize=1\linewidth
    \includegraphics[width=9cm]{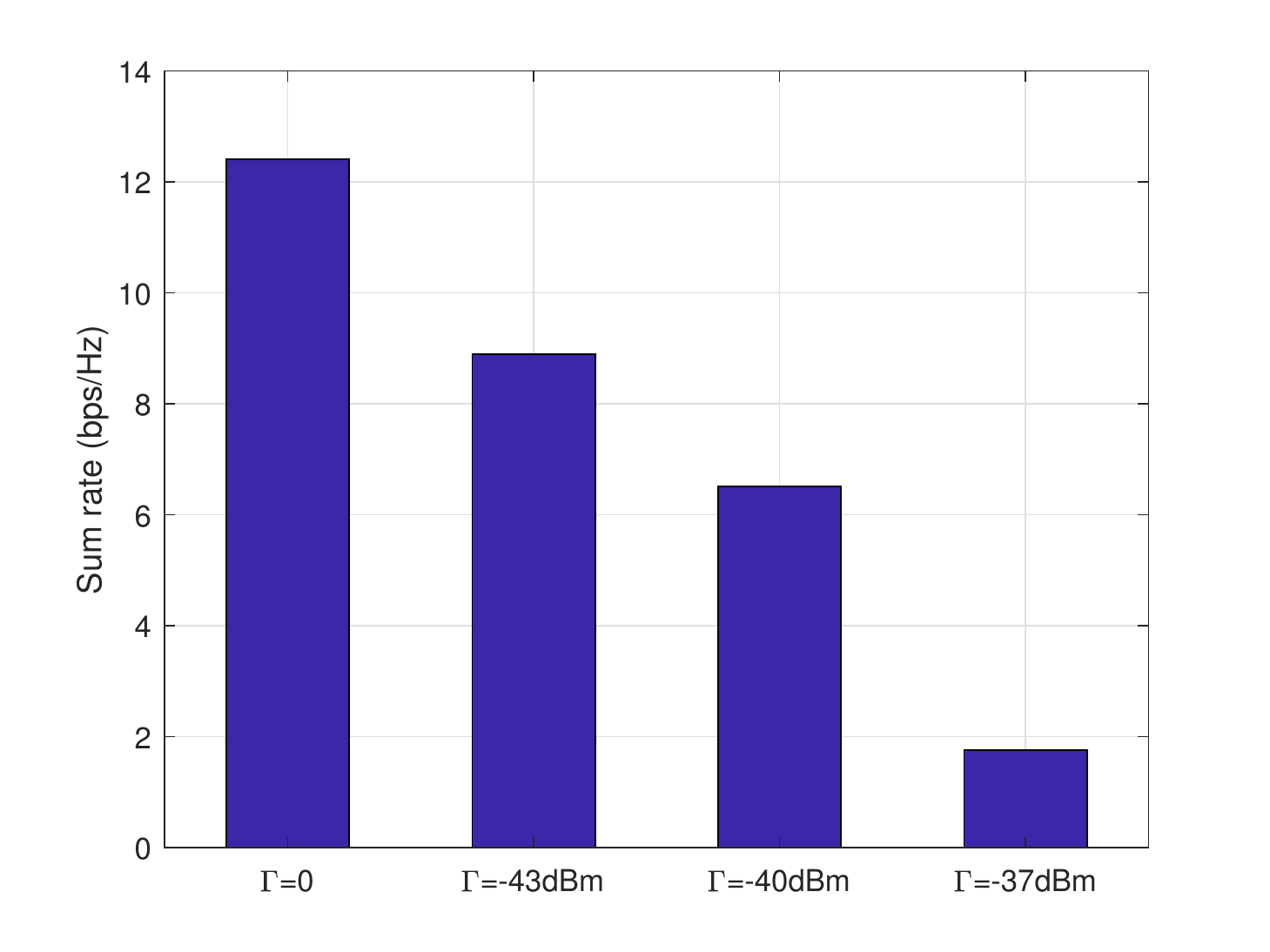}
\vspace{-15pt}
\caption{The communication sum rate versus the sensing beampattern threshold $\Gamma$.}
\label{hover sum rate vers gamma}
\vspace{-10pt}
\end{figure}

Fig. \ref{simulation setup} and Fig. \ref{hover sum rate vers gamma} show the obtained UAV deployment locations and the correspondingly achieved sum rates under different values of sensing beampattern threshold $\Gamma$. First, it is observed that two symmetric UAV deployment locations are obtained under each realization with different $\Gamma$. This is intuitive, as the sensing area and communication users are deployed in a  symmetric manner. It is also observed that as $\Gamma$ increases (e.g., from $0$ to  $-37~{\rm dBm}$), the UAV is deployed closer to the sensing area, but further away from the communication users, thus leading to decreased communication sum rate. %This is due to the fact that in this case, the UAV needs to increases the beampattern gain at interested areas by not only reducing the physical distance but also adapting the beam directions towards them, thus leading to reduced communication rates.

\begin{figure}[h]
\centering
\subfigure[Sensing only.]{
\begin{minipage}[t]{0.315\linewidth}
\centering
\includegraphics[width=6cm]{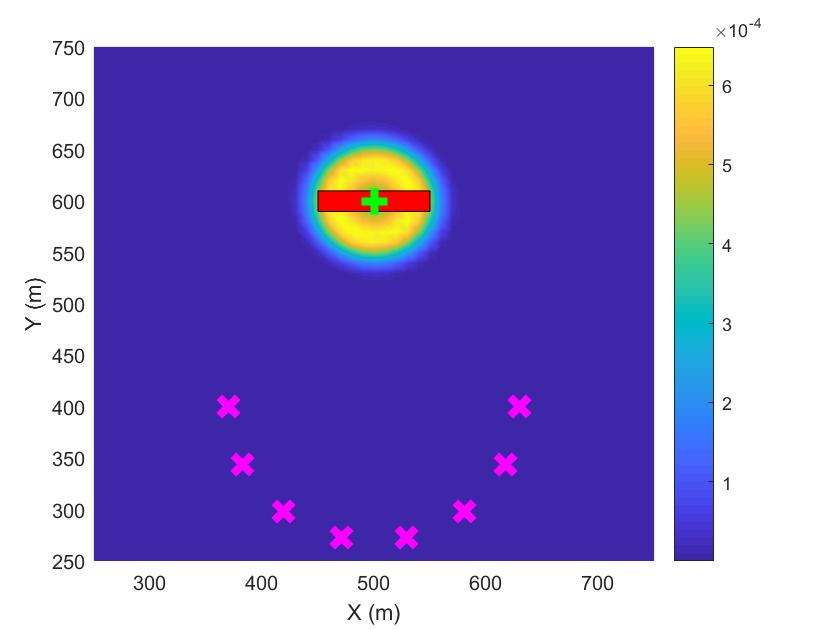}
%\caption{fig2}
\end{minipage}
\label{hover beampattern sen}
}%
\subfigure[Communication only.]{
\begin{minipage}[t]{0.315\linewidth}
\centering
\includegraphics[width=6cm]{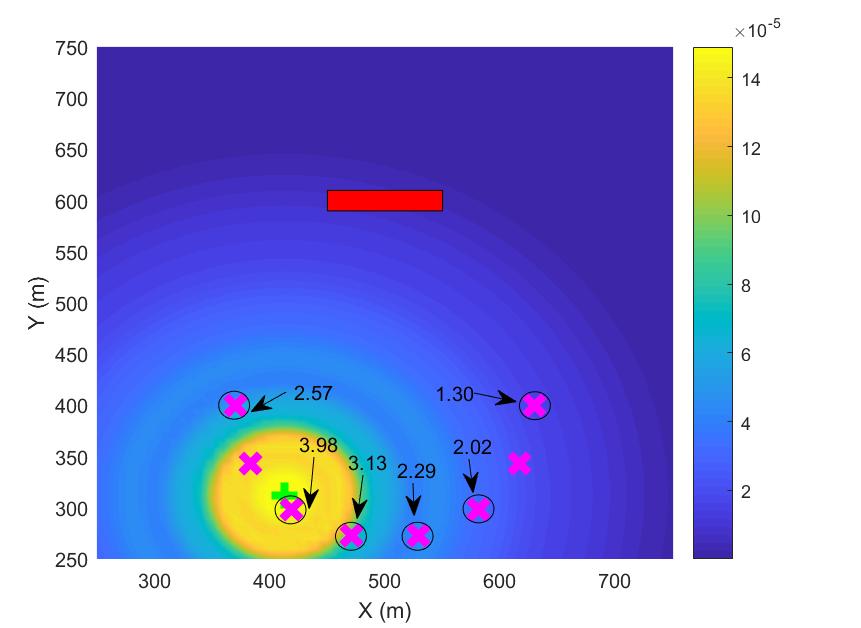}
%\caption{fig1}
\end{minipage}%
\label{hover beampattern com}
}%
\subfigure[Proposed ISAC design.]{
\begin{minipage}[t]{0.315\linewidth}
\centering
\includegraphics[width=6cm]{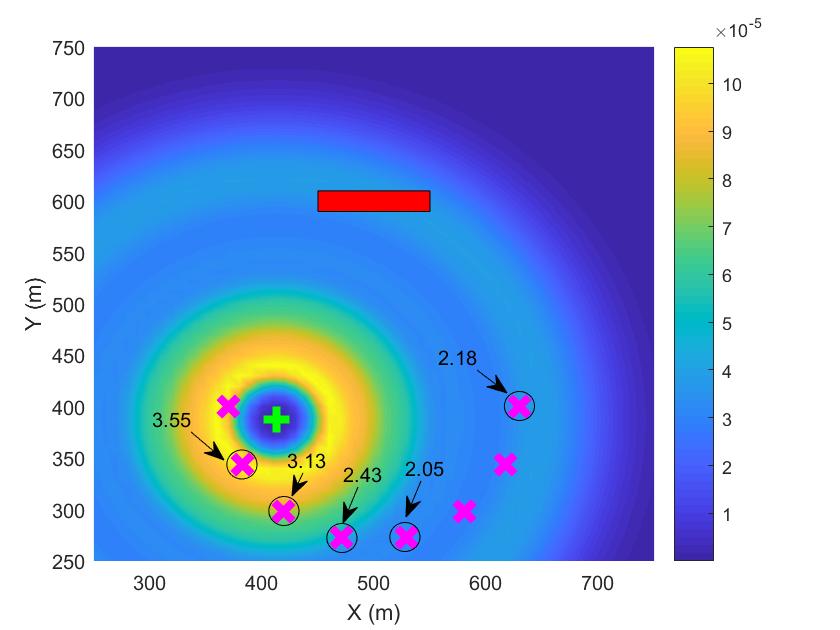}
%\caption{fig1}
\end{minipage}%
\label{hover beampattern ISAC}
}%
\centering
\caption{Obtained deployment locations and corresponding beampattern gains in space under different designs, where the red rectangle denotes the sensing area, the green '$+$' denotes the obtained deployment location of the UAV,  the carmine $\times$ denotes the location of each communication user, and the number associated with each user corresponds to its communication rate (in bps/Hz) if served.}
\label{hover beampattern}
\end{figure}

Fig. \ref{hover beampattern} shows the obtained deployment locations of UAV and the corresponding (receive) beampattern gains in space under different designs with $\Gamma = -45~{\rm dBm}$ and $M=16$. First, for sensing design in Fig. \ref{hover beampattern sen}, the UAV is observed to be deployed at the center of the sensing area, and the sensing power exactly covers the whole sensing area, thanks to the properly designed sensing beams in this case. Next, for the communication only design in \ref{hover beampattern com}, it is observed that the UAV is deployed above the communication users, and the UAV's transmission power is radiated towards users in order to efficiently perform the task of communication. Finally, for our proposed ISAC design in Fig. \ref{hover beampattern ISAC}, it is observed that the UAV is deployed between the users and the sensing area. Moreover, the beampattern gains in Fig. \ref{hover beampattern ISAC} are observed to be more uniformly distributed as compared to those in Fig. \ref{hover beampattern sen} and Fig. \ref{hover beampattern com} to well balance the tradeoff between communication and sensing performances.

 %For ISAC with ZF beamforming, the inter-user interference is eliminated and the radar signal is projected into the null space of the channel matrix, which limits the DoF of the information and sensing beamforming design, and thus compromises the communication performance. 

%\begin{figure}[h]
%\centering
%\subfigure[Simulation setup and the obtained deployment locations of UAV.]{
%\begin{minipage}[t]{0.5\linewidth}
%\centering
%\includegraphics[width=8cm]{hover_gamma.eps}
%%\caption{fig1}
%\end{minipage}%
%}%
%\subfigure[Communication rates under different $\Gamma$.]{
%\begin{minipage}[t]{0.5\linewidth}
%\centering
%\includegraphics[width=8cm]{hover_rate.eps}
%%\caption{fig2}
%\end{minipage}%
%}%
%\centering
%\caption{Optimal hovering locations and corresponding communication rates for ISAC under different $\Gamma$.}
%\end{figure}

\vspace{-10pt}

\subsection{Mobile UAV Scenario}
\begin{figure}[h]
		\centering
		\includegraphics[width=9cm]{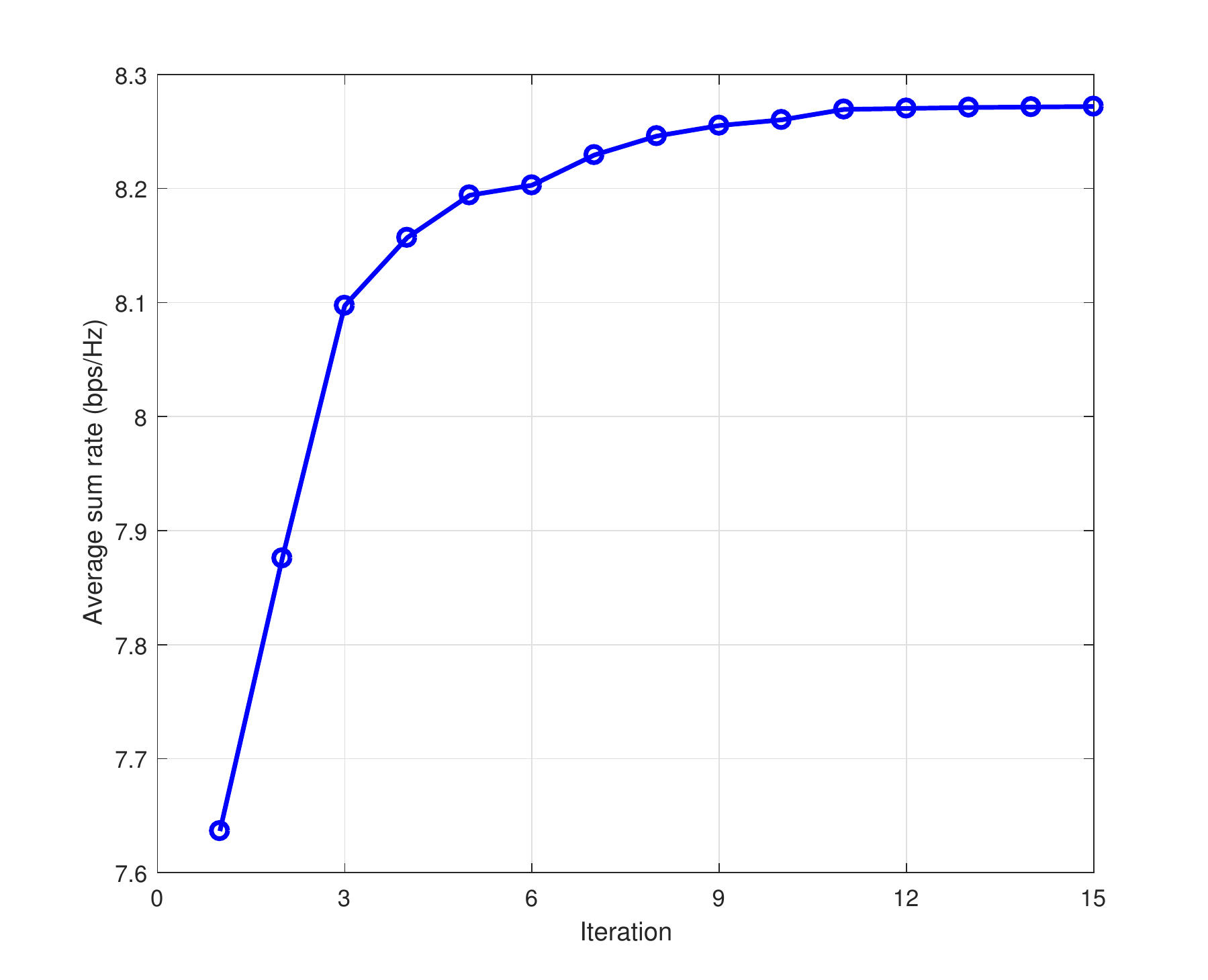}
\vspace{-15pt}
		\caption{Convergence behavior of the proposed trust-region-based SCA algorithm for solving problem (P2).}
\vspace{-3pt}
\label{converge}
\end{figure}

Next, we consider the mobile UAV scenario. Before the performance comparison, we first show the convergence behavior of the proposed algorithm in Fig. \ref{converge}. It can be observed that the average sum rate achieved by our proposed design increases quickly with the number of iterations and the algorithm converges in about 12 iterations.
\vspace{-17pt}

Then the following benchmark schemes are considered for performance comparison. 
\begin{itemize}
\item \textbf{Straight flight (SF)}: The UAV flies from the initial location $\hat{ \boldsymbol{ q}}^{\rm I} $ to the final location $\hat{ \boldsymbol{ q}}^{\rm F}$ straightly by using the constant speed of $\|\hat{ \boldsymbol{ q}}^{\rm F}-\hat{ \boldsymbol{ q}}^{\rm I}\|/T$.  Accordingly, we optimize the information and sensing beamforming $\{{\boldsymbol w}_k[n]\}$ and $\{{\boldsymbol R}_s[n]\} $ via solving problem (P6) under such predetermined SF trajectory.
 \item \textbf{Fly-hover-fly (FHF)}: The UAV first flies straightly from the initial location $\hat{ \boldsymbol{ q}}^{\rm I} $ to the optimized hovering location obtained from problem (P3) at the maximum speed ${\tilde V}_{\max}$, then hovers at this point, and finally flies straightly towards the final location $\hat{ \boldsymbol{ q}}^{\rm F}$ also at speed ${\tilde V}_{\max}$. Under such FHF trajectory, we optimize $\{{\boldsymbol w}_k[n]\}$ and $\{{\boldsymbol R}_s[n]\} $ via solving problem (P6).

%\quad For both SF and FHF trajectories, it may happen that some locations along $\{ {\boldsymbol{\dot q}}[n]\} $ may be infeasible, so we need to recast the above trajectory to the feasible region ${{\mathcal{Q}}^{\rm h}}$ obtained in section II by solving the following problem.
%\begin{align}
%\text{(P11)}:\mathop {\min }\limits_{\{ {\boldsymbol{\bar q}}[n]\}} ~& \|{\boldsymbol{\dot q}}[n] - {\boldsymbol{\bar q}}[n]\|{^2}   \nonumber\\
%\mathrm{s.t.}~&
%\left\| {{\boldsymbol{\bar q}}[n + 1] - {\boldsymbol{\bar q}}[n]} \right\| \le {V_{\max }}{\Delta _t}, \forall n \in \mathcal N\tag{43a}\\
%~
%&{\boldsymbol{\bar q}}[n] \in {{{\mathcal Q}}^{\rm h}},\forall n \in \mathcal N. \tag{43b}
%\end{align}
%Finally, the fixed trajectory $\{ {\boldsymbol{\bar q}}[n]\}$ is obtained.
%\item \textbf{Zero-forcing (ZF) beamforming with SF}: In this case, under the above obtained SF trajectory, we optimize $\{{\boldsymbol w}_k[n]\}$ and $\{{\boldsymbol R}_s[n]\} $ based on zero-forcing
%the inter-user interference and radar interference, which is equivalent to solving (P9-ZF) for each slot $n$.
%    \item \textbf{Zero-forcing (ZF) beamforming with FHF}: In this case, we design $\{{\boldsymbol w}_k[n]\}$ and $\{{\boldsymbol R}_s[n]\} $ via zero-forcing beamforming technique under the obtained FHF trajectory, which is equivalent to solving (P9-ZF) for each slot $n$.
\item \textbf{Communication only}: The UAV only performs the communication task. This corresponds to solving problem (P2) by setting $\Gamma=0$.
    \item \textbf{Sensing only}: The UAV only performs the sensing task. In this case, we aim to optimize the UAV's trajectory jointly with the sensing beamforming to maximize the weighted minimum beampattern gain, for which the optimization problem is formulated as
\begin{align}
\text{(P10)}:\mathop {\max }\limits_{\{ {{\boldsymbol R}_s[n]}\succeq 0, \atop {\boldsymbol q}[n]\}} & {\min_{j\in\mathcal J}} ~ \frac{1}{d^2({\boldsymbol q}[n],{{\boldsymbol m}_j})}{\boldsymbol{a}}^{\rm H}({\boldsymbol{{q}}}[n],{{\boldsymbol{m}}_j}){{\boldsymbol{R}}_s} [n]{\boldsymbol{a}}({\boldsymbol{{q}}}[n],{{\boldsymbol{m}}_j}) \nonumber\\
\mathrm{s.t.} ~&{\rm{tr}}({{\boldsymbol{R}}_s}[n]) \le {P_{\max }}, \forall n \in \mathcal N   \label{P10 power}\\
~
&(11),~(12),~\text{and}~(13). \nonumber
\end{align}
Problem (P10) can be solved similarly as problem (P2), for which the details are omitted for brevity.
\end{itemize}

\begin{figure}[h]
		\centering
		\includegraphics[width=9cm]{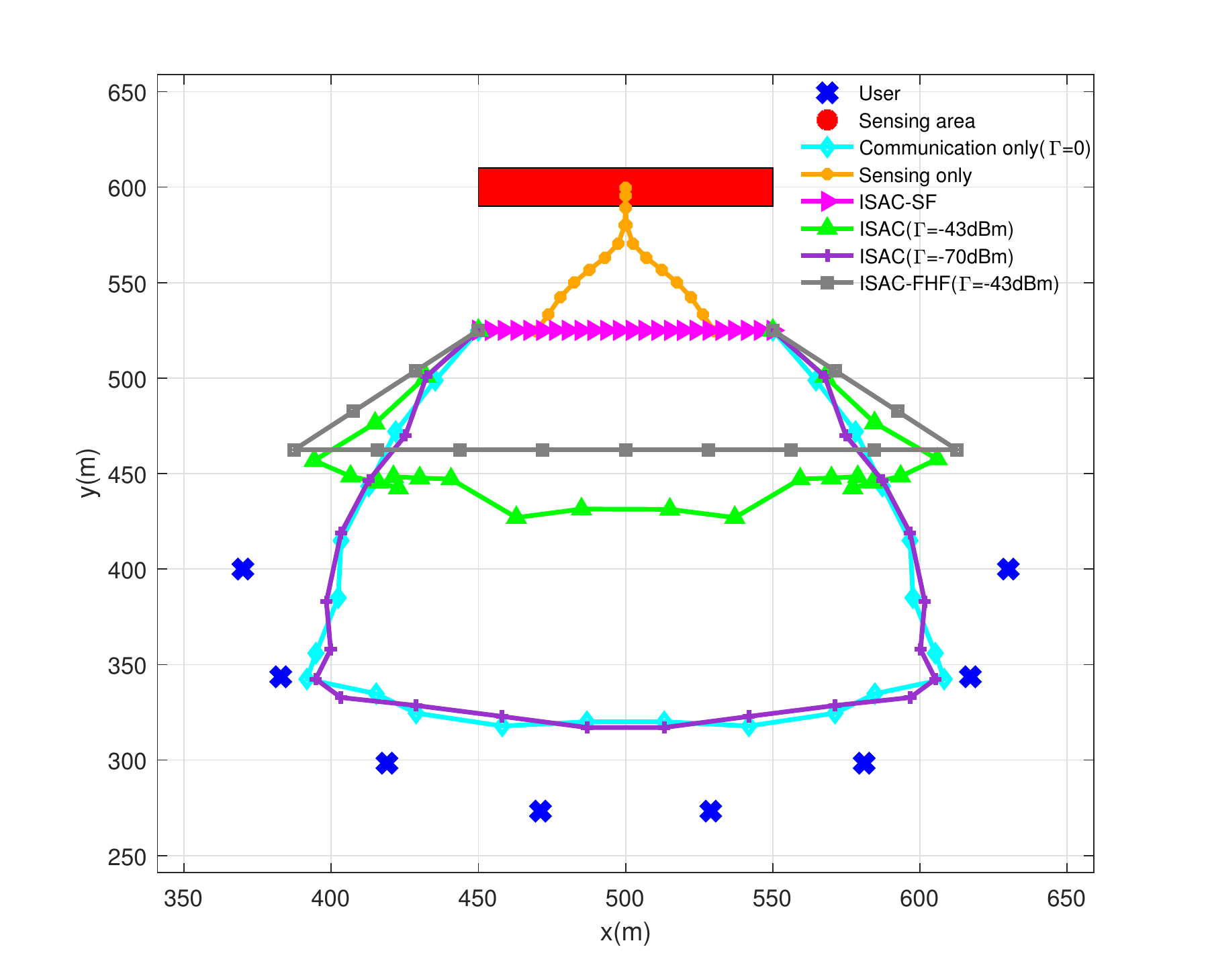}
\vspace{-15pt}
		\caption{Obtained trajectories under different designs.}
\label{traj}
\end{figure}

Fig. \ref{traj} shows the obtained UAV flight trajectories under different designs. For the design with sensing only, it is observed that the UAV flies towards the center of the sensing area to maximize the beampattern gains therein. For our proposed ISAC designs, it is observed that as $\Gamma$ reduces from $-43~{\rm dBm}$ to $0$, the UAV flies much closer to the  users to increase the communication rate while satisfying the sensing requirements. More specifically, when $\Gamma$ is sufficiently small (i.e., $\Gamma = -70~{\rm dBm}$), the obtained UAV trajectory is observed to be similar as that with communication only. This shows that in this case, the sensing requirements can be easily ensured by reusing the information signals that are dedicatedly designed for rate maximization.
\vspace{-10pt}

\begin{figure}[h]
\centering
\subfigure[$n=1$]{
\begin{minipage}[t]{0.315\linewidth}
\centering
\includegraphics[width=5.8cm]{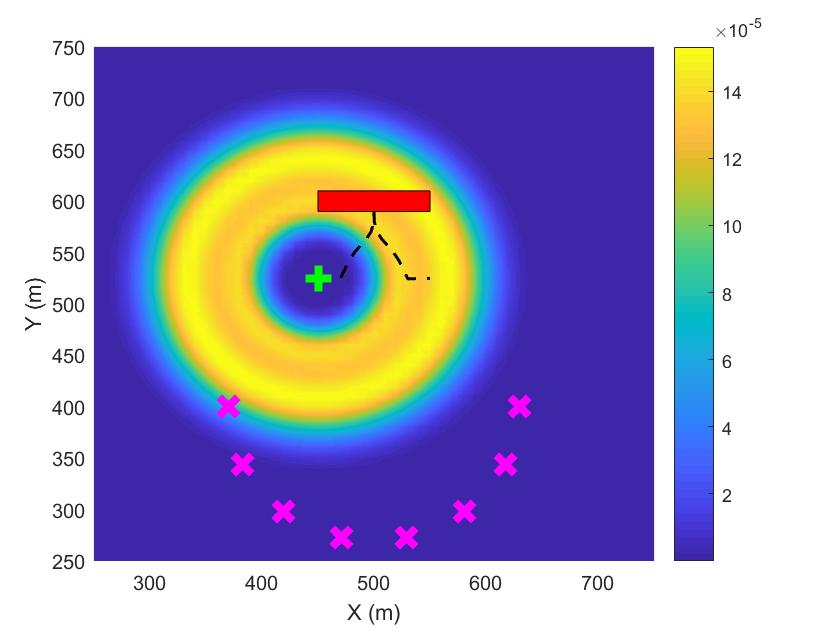}
%\caption{fig1}
\end{minipage}%
}%
\subfigure[$n=6$]{
\begin{minipage}[t]{0.315\linewidth}
\centering
\includegraphics[width=5.8cm]{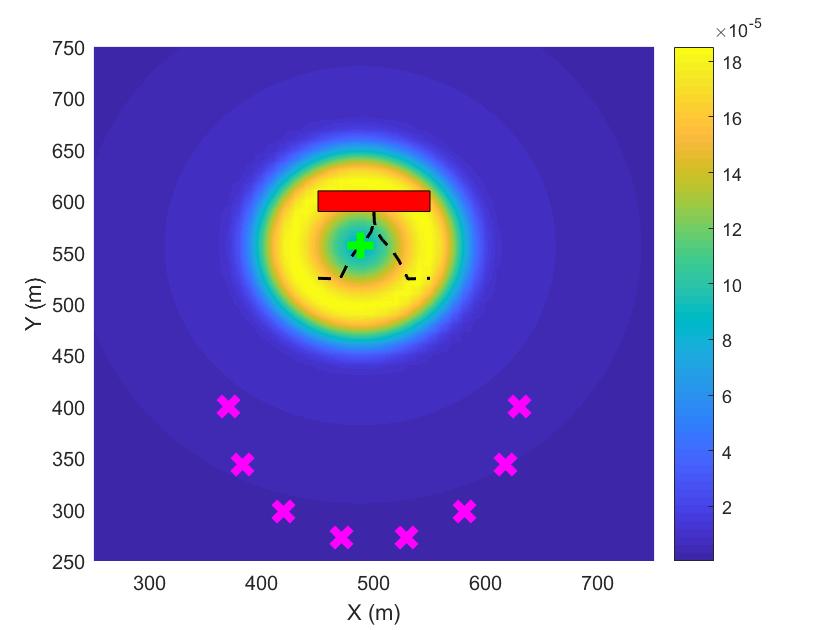}
%\caption{fig2}
\end{minipage}%
}%
\subfigure[$n=12$]{
\begin{minipage}[t]{0.315\linewidth}
\centering
\includegraphics[width=5.8cm]{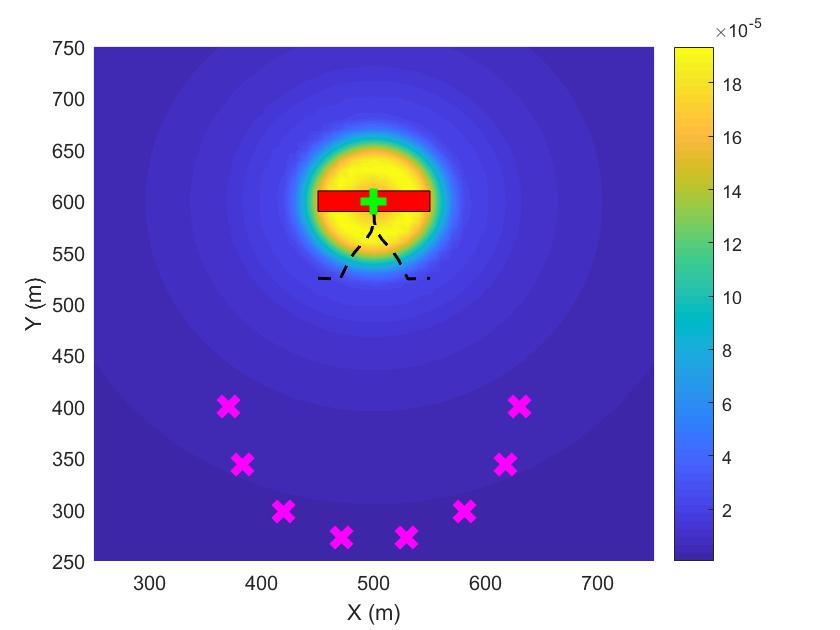}
%\caption{fig2}
\end{minipage}
}%
\centering
\caption{Achieved beampattern gains in space under the sensing only design at specifically chosen time slots $n=1,6,12$, respectively, where the red rectangle denotes the sensing area, the black dashes line denotes the UAV's flight trajectory, the green '$+$' denotes the UAV's location at the specified time slot, and the carmine '$\times$' denotes the location of each user.}
\label{sen-only}
\end{figure}

\begin{figure}[h]
\centering
\subfigure[$n=1$]{
\begin{minipage}[t]{0.315\linewidth}
\centering
\includegraphics[width=5.8cm]{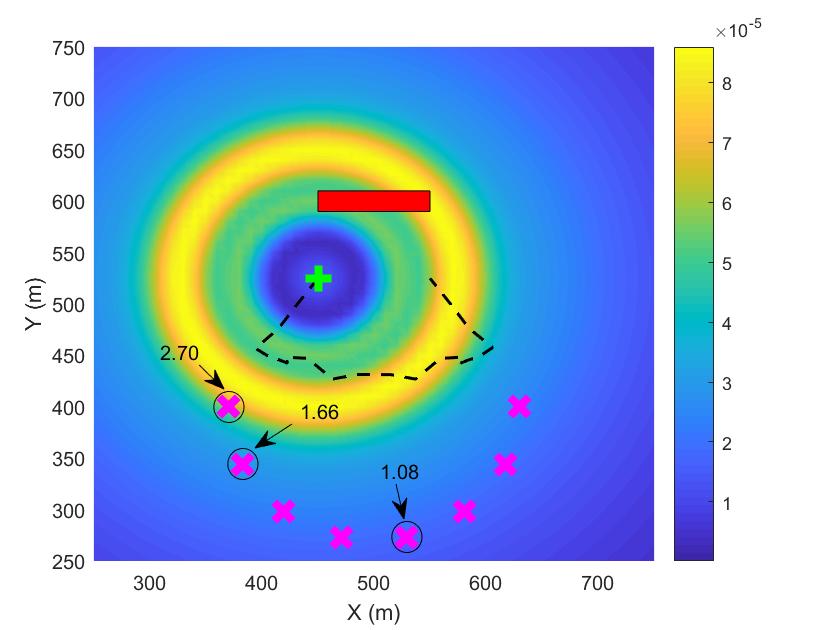}
%\caption{fig1}
\end{minipage}%
}%
\subfigure[$n=6$]{
\begin{minipage}[t]{0.315\linewidth}
\centering
\includegraphics[width=5.8cm]{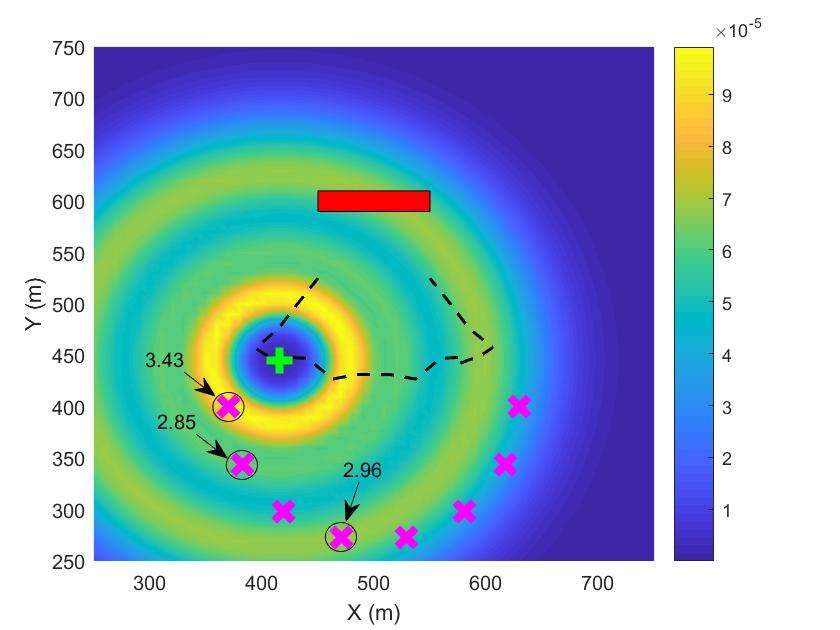}
%\caption{fig2}
\end{minipage}%
}%
\subfigure[$n=12$]{
\begin{minipage}[t]{0.315\linewidth}
\centering
\includegraphics[width=5.8cm]{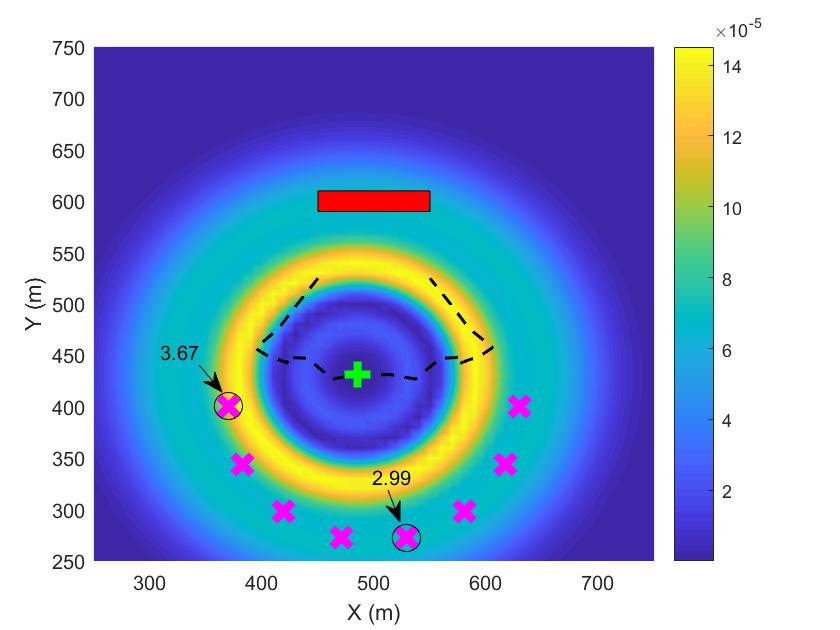}
%\caption{fig2}
\end{minipage}
}%
\centering
\caption{Achieved beampattern gains in space under the proposed ISAC design at specifically chosen time slots $n=1,6,12$, respectively, with the beampattern gain threshold $\Gamma=-43~{\rm dBm}$, where the red rectangle denotes the sensing area, the black dashes line denotes the UAV's flight trajectory, the green '$+$' denotes the UAV's location at the specified time slot, the carmine '$\times$' denotes the location of each user, and the number associated with each user corresponds to its communication rate (in bps/Hz) if served.}
\label{ISAC}
\end{figure}

\begin{figure}[h]
\centering
\subfigure[$n=1$]{
\begin{minipage}[t]{0.315\linewidth}
\centering
\includegraphics[width=5.8cm]{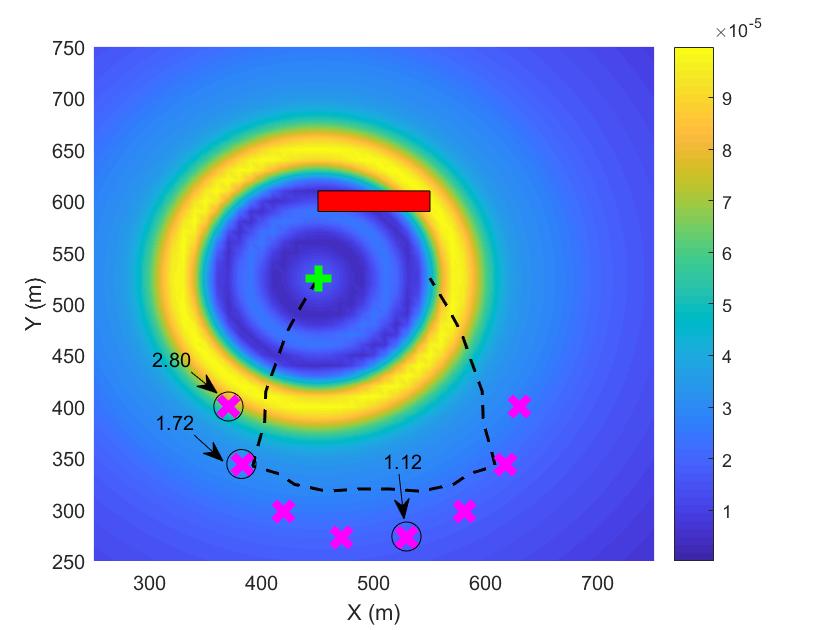}
%\caption{fig1}
\end{minipage}%
}%
\subfigure[$n=8$]{
\begin{minipage}[t]{0.315\linewidth}
\centering
\includegraphics[width=5.8cm]{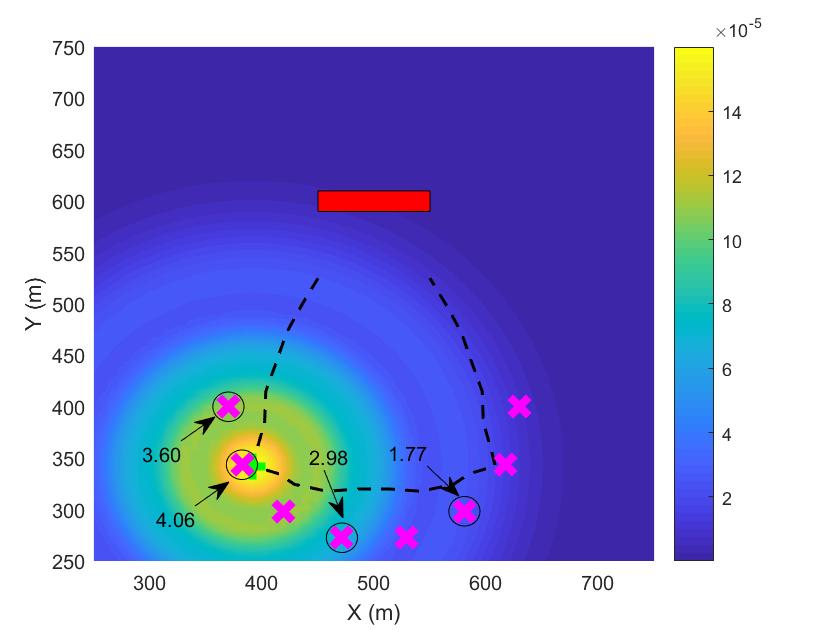}
%\caption{fig2}
\end{minipage}%
}%
\subfigure[$n=12$]{
\begin{minipage}[t]{0.315\linewidth}
\centering
\includegraphics[width=5.8cm]{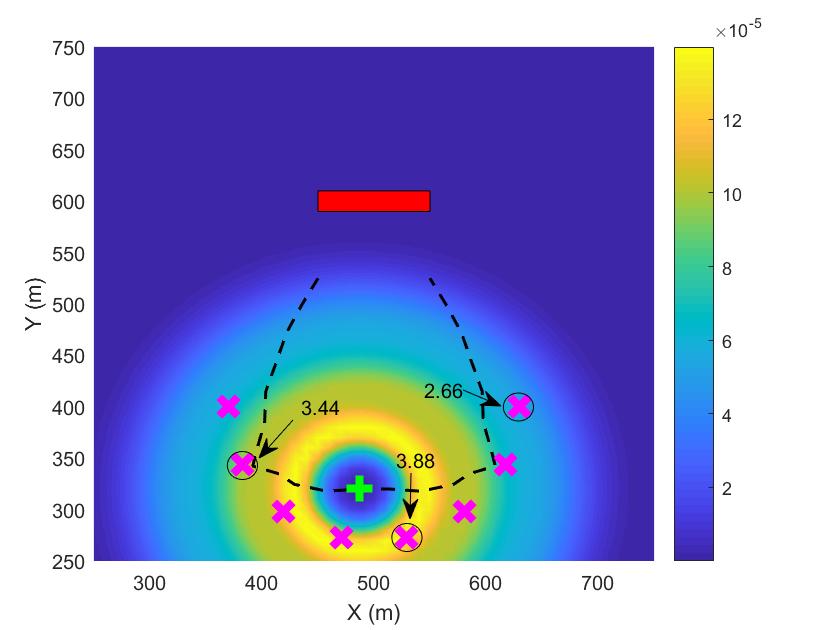}
%\caption{fig2}
\end{minipage}
}%
\centering
\caption{Achieved beampattern gains in space under the communication only design at specifically chosen time slots $n=1,8,12$, respectively, where the red rectangle denotes the sensing area, the black dashes line denotes the UAV's flight trajectory, the green '$+$' denotes the UAV's location at the specified time slot, the carmine '$\times$' denotes the location of each user, and the number associated with each user corresponds to its communication rate (in bps/Hz) if served.}
\label{com-only}
\end{figure}

Figs. \ref{sen-only}, \ref{ISAC}, and \ref{com-only} show the obtained (receive) beampattern gains in space at specifically chosen time slots for the sensing only design, our proposed ISAC design, and the communication only design, respectively. First, for the sensing only and communication only designs in Figs. \ref{sen-only} and \ref{com-only}, the similar observations can be made as those from Figs. \ref{hover beampattern sen} and \ref{hover beampattern com} in the quasi-stationary UAV scenario. Next, for our proposed ISAC design in Fig. \ref{ISAC}, it is observed that the transmission energy is properly radiated to circles with different radius for covering both sensing areas and communication users, thus balancing their performance tradeoff.

% When designing the ISAC system, we should consider a reasonable beampattern gain threshold $\Gamma$ to balance the tradeoff between the sensing and communication functions. 

%\footnote{Note that the location of the UAV changes over time, thus the AoDs between the UAV and the users (as well as the sensing targets) also change over time. Also, from the above figures, it can be observed that the obtained trajectories are symmetric. Without loss of generality, we simply choose certain time slots during the first half of the mission duration for presentation. } 

It is also interesting to discuss the instantaneous communication rates achieved by different users in Figs. \ref{ISAC} and \ref{com-only}. It is observed that the users with distinct distances with the UAV are likely to be served simultaneously to achieve optimized sum rate. This is due to the fact that with vertically deployed ULA at the UAV, the users at distinct distances with the UAV would have more diverse AoDs with each other, thus resulting in less co-channel interference that is beneficial for the multiuser MIMO communication.

%the farthest and the nearest users are more likely to be served by the UAV, because such design may alleviate the co-channel interference.

%Furthermore, in , we label the instaneous communication rate of the users being served. In general, the farthest and the nearest users are more likely to be served by the UAV, because such design may alleviate the co-channel interference. However, for the farthest user, there exist a tradeoff between its own signal propagation loss and its interference to other users. Specifically, although with such design, it may impose little interference on others, its signal power may be attenuated greatly by the large path loss, which will lead to low communication rate. Thus, sometimes simply choosing the farthest user to serve is not beneficial to enhance the overall sum-rate throughput. In this case, the UAV may consider such tradeoff and serve other users (such as the second farthest) instead (as shown in Fig. 7 (b) and Fig. 8(b)).

\begin{figure}
	\centering
	\begin{minipage}{.45\linewidth}
		\centering
		\includegraphics[width=8.2cm]{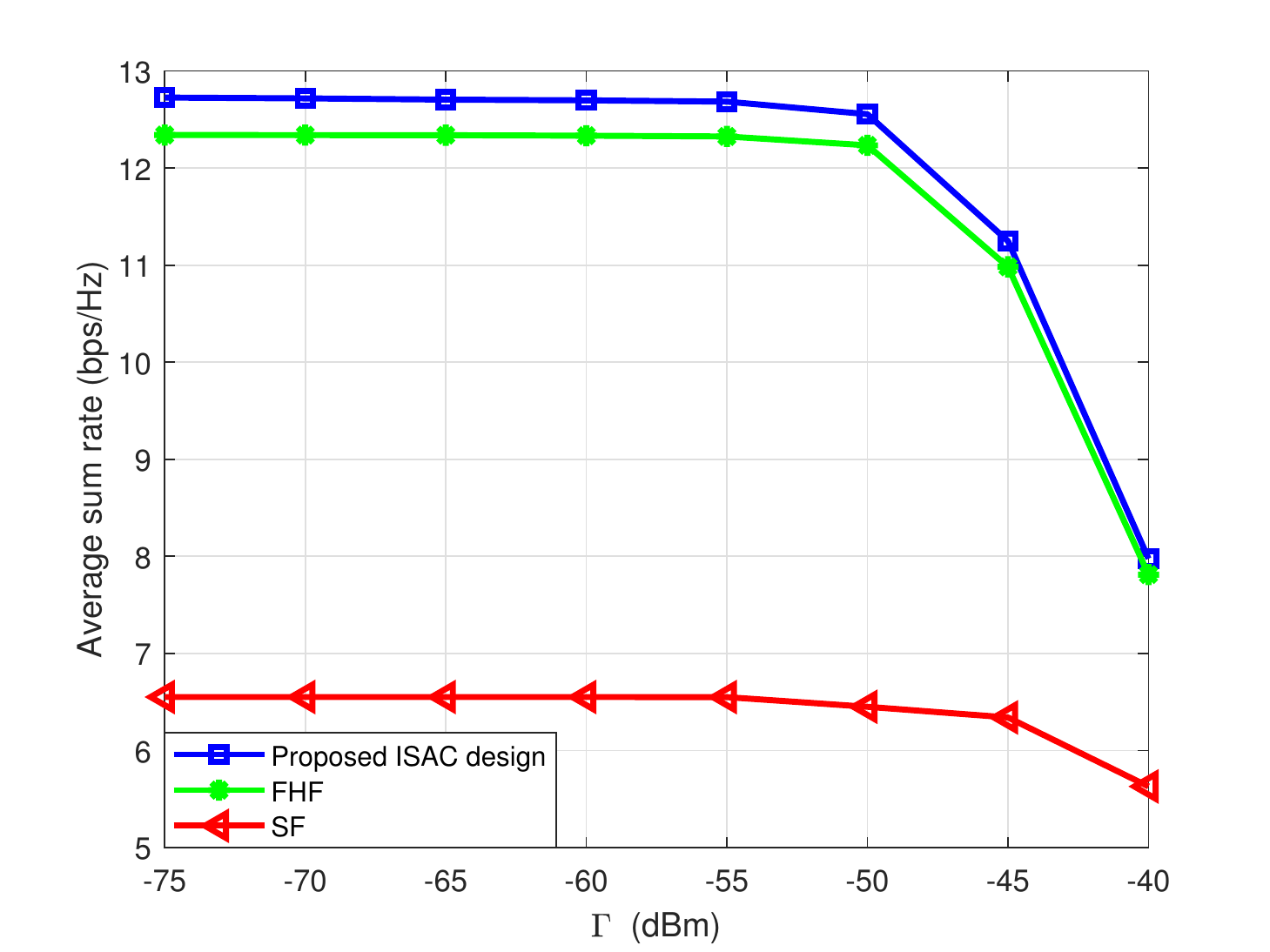}
		\caption{The average sum rate versus the beampattern gain threshold $\Gamma$.}
\label{traj-rate-gamma}
	\end{minipage}
	\begin{minipage}{.45\linewidth}
		\centering
		\includegraphics[width=8.2cm]{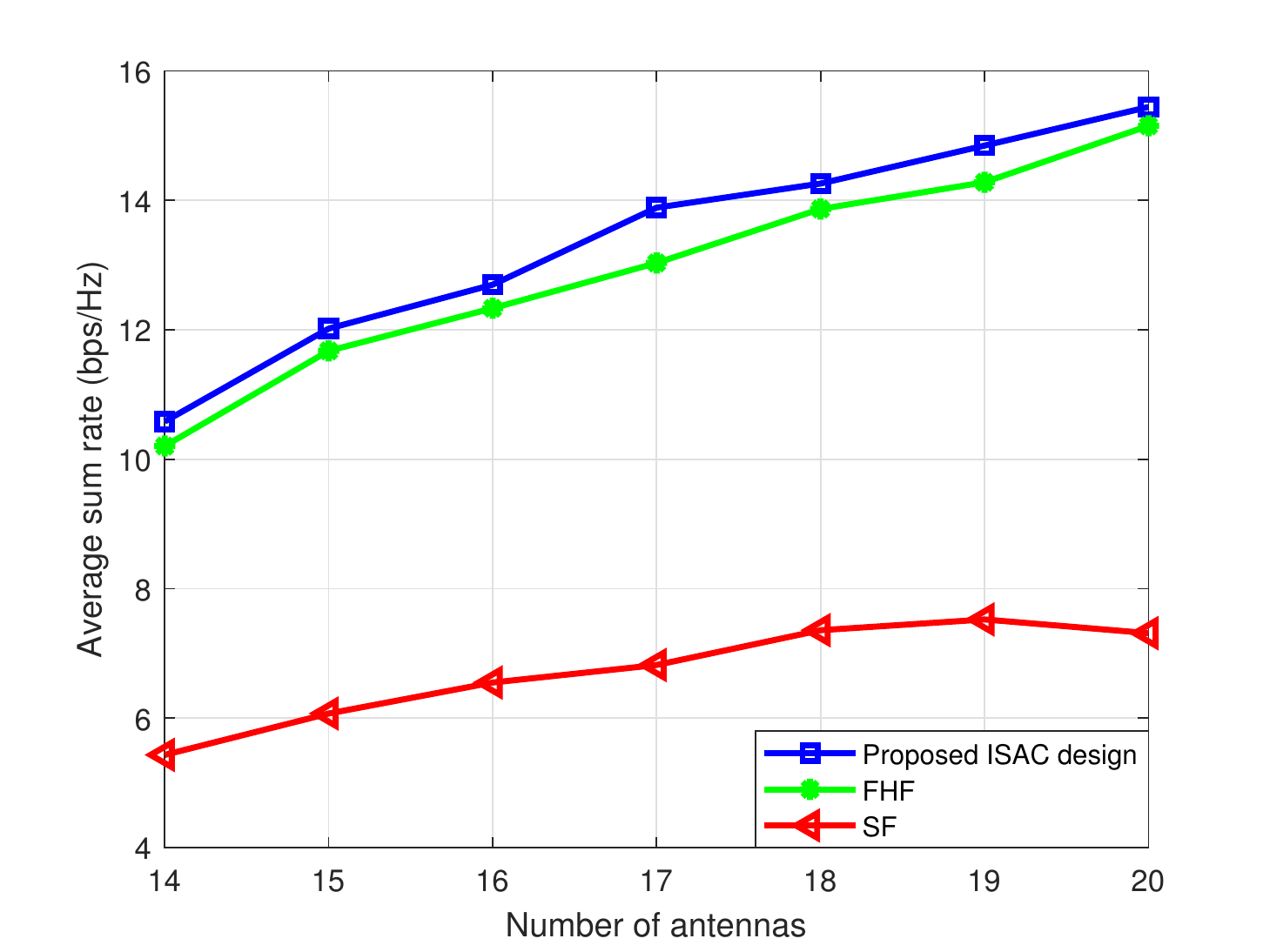}
		\caption{The average sum rate  versus the number of transmit antennas $M$ at UAV.}
\label{traj-rate-M}
	\end{minipage}
\end{figure}

Fig. \ref{traj-rate-gamma} shows the average sum rate at users versus the beampattern gain threshold $\Gamma$ with $M=16$. It is observed that as $\Gamma$ increases, the average sum rate decreases for all the three schemes, as the UAV needs to spend more transmission power for taking care of sensing. It is also observed that the proposed ISAC design and the FHF design significantly outperform the SF design. This shows the benefit of our proposed optimization for UAV trajectory and deployment (hovering) location. As $\Gamma$ becomes large, it is observed that the performance gap among three designs significantly reduces. This is due to the fact that in this case, the feasible flight region for ensuring the sensing requirement would become limited, thus limiting the design DoF in trajectory optimization.

%as $\Gamma$ decreases, the sum-rate throughput gradually converges to a constant value. This is because when $\Gamma$ is smaller than a threshold, the feasible region will cover all the mission area, thus the optimal hovering locations (trajectory) as well as the information and sensing beamforming vectors remain the same. Furthermore, it can be easily observed that the joint design of trajectory, information beamforming vectors, and sensing covariance matrix outperforms other benchmark schemes significantly. This is because the joint design method can offer additional design DoF in terms of improving the system performance compared to other benchmark schemes. 

%Finally, under both of the benchmark trajectory design schemes (i.e. FHF and SF), our proposed OB-based design via solving (P6) outperforms ZF-based design. This is because, in ZF beamforming, the limited design DoF of the information and radar beamforming may introduce performance loss compared to optimized beamforming in our proposed design (or even makes the problem infeasible under some setups).

Fig. \ref{traj-rate-M} shows the average sum rate versus the number of antennas $M$ at the UAV, where we set $\Gamma= -60~{\rm dBm}$. It is observed that our proposed ISAC design and the FHF design outperform the SF design. More specifically, when $M$ becomes large, the average sum rates by our proposed ISAC design and the FHF design are observed to increase almost in a linear manner, while that by the SF design is observed to saturate. This is due to the fact that with proper trajectory or hovering location optimization, the UAV can enjoy narrower beams provided by more antennas to increase the information signal power while better mitigating the inter-user interference.

%Moreover, it can be also observed that a higher average sum-rate throughput is generally achieved with a larger $M$ (e.g., from 14 to 20). This is because the UAV can further exploit spatial multiplexing and diversity gains with more antennas, which can increase the design DoF of the UAV-enabled ISAC systems for performance enhancement. For SF with OB, when $M$ is larger than 19, the sum-rate throughput is slightly decreased. This is probably because increasing $M$ may narrow the beam of the directional antenna, which may compromise the system performance.

\section{Conclusion}
This paper considered a novel UAV-enabled ISAC system, where the UAV acts as an aerial dual-functional AP to simultaneously communicate with multiple users and perform radar sensing towards an interested area. We exploited the UAV's mobility to improve the communication data rate throughput while ensuring the sensing requirements. Under two scenarios with quasi-stationary and mobile UAVs, we  designed the UAV's placement and trajectory jointly with the transmit information and sensing beamforming. We proposed efficient algorithms to solve the highly non-convex sensing-constrained rate maximization problems, by using techniques from convex and non-convex optimization. Numerical results showed that our proposed designs significantly outperform other benchmark schemes. How to extend the results to other scenarios (e.g., with other types of antenna configurations at the UAV and with multiple or swarm UAVs) is interesting and worth pursuing in future work.

\begin{appendix}
\subsection{Proof of Proposition \ref{rankone}}
For notational convenience, we omit the superscript $(l)$ in the following. Suppose that $\{ {{\boldsymbol{\dot W}}_k}\}$ and ${{\boldsymbol{\dot R}}_s} $ denote the obtained optimal solution to problem (SDR5.$l$). Based on them, we then construct a new solution $\{ {{\boldsymbol{\bar W}}_k}\} $ and ${{\boldsymbol{\bar R}}_s}$ (and the corresponding $\{{{\boldsymbol{\bar w}}_k}\}$) as
\begin{align}\label{w rankone}
{{\boldsymbol{\bar w}}_k} = {({{\boldsymbol{h}}_k^{\rm{H}}({\boldsymbol{q}},{{\boldsymbol{u}}_k})}{{\boldsymbol{\dot W}}_k}{{\boldsymbol{h}}_k({\boldsymbol{q}},{{\boldsymbol{u}}_k})})^{ - 1/2}}{{\boldsymbol{\dot W}}_k}{{\boldsymbol{h}}_k({\boldsymbol{q}},{{\boldsymbol{u}}_k})},
\end{align}
\begin{align}\label{W rankone}
{{\boldsymbol{\bar W}}_k}= {{\boldsymbol{\bar w}}_k}{\boldsymbol{\bar w}}_k^{\rm H}, 
\end{align}
\begin{align}\label{r rankone}
{{\boldsymbol{\bar R}}_s} = \sum\nolimits_{k = 1}^K {{{{\boldsymbol{\dot W}}}_k}}  + {{\boldsymbol{\dot R}}_s} - \sum\nolimits_{k = 1}^K {{{{\boldsymbol{\bar W}}}_k}}.
\end{align}
In the following, we show that $\{ {{\boldsymbol{\bar W}}_k}\}$ and ${{\boldsymbol{\bar R}}_s}$  are optimal for problem (SDR5.$l$). It is evident that $\{{\boldsymbol{\bar W}}_k\}$ are positive semidefinite and rank-one. Also, according to \eqref{r rankone}, the constraints for beampattern gain in \eqref{P4 beampattern} and those in transmit power in \eqref{P4 power} hold for $ \{ {{\boldsymbol{\bar W}}_k}\}$ and ${{\boldsymbol{\bar R}}_s}$. 

Next, we show that ${{\boldsymbol{\bar R}}_s}$ is positive semidefinite. For any ${\boldsymbol r} \in \mathbb{C}^{M \times 1}$, it holds that 
\begin{align}
{{\boldsymbol{r}}^{\rm H}}\left( {{{{\boldsymbol{\dot W}}}_k}\! - \!{{{\boldsymbol{\bar W}}}_k}} \right){\boldsymbol{r}} = {{\boldsymbol{r}}^{\rm H}}{{\boldsymbol{\dot W}}_k}{\boldsymbol{r}} \!-\! {\left( {{{\boldsymbol{h}}_k^{\rm{H}}({\boldsymbol{q}},{{\boldsymbol{u}}_k})}{{{\boldsymbol{\dot W}}}_k}{{\boldsymbol{h}}_k({\boldsymbol{q}},{{\boldsymbol{u}}_k})}} \right)^{ - 1}}{\left| {{\boldsymbol{r}}^{\rm H}{{{\boldsymbol{\dot W}}}_k}{{\boldsymbol{h}}_k({\boldsymbol{q}},{{\boldsymbol{u}}_k})}} \right|^2},\forall k \in \mathcal{K}. 
\end{align}
According to the Cauchy-Schwarz inequality, we have
\begin{align}
{\left| {{{\boldsymbol{r}}^{\rm H}}{{{\boldsymbol{\dot W}}}_k}{{\boldsymbol{h}}_k({\boldsymbol{q}},{{\boldsymbol{u}}_k})}} \right|^2}&={\left| {{{\boldsymbol{r}}^{\rm H}}{{\boldsymbol{\dot w}}_k}{\boldsymbol{\dot w}}_k^{\rm H}{{\boldsymbol{h}}_k({\boldsymbol{q}},{{\boldsymbol{u}}_k})}} \right|^2} \nonumber \\
&\le {\left| {{{\boldsymbol{r}}^{\rm H}}{{\boldsymbol{\dot w}}_k}} \right|^2} {\left| {{{{\boldsymbol{h}}^{\rm H}_k({\boldsymbol{q}},{{\boldsymbol{u}}_k})}}{{\boldsymbol{\dot w}}_k}} \right|^2}\nonumber\\
%& =( {{{\boldsymbol{r}}^{\rm H}}{{\boldsymbol{\dot w}}_k}{\boldsymbol{\dot w}}_k^{\rm H}{\boldsymbol{r}}} )\left( {{{\boldsymbol{h}}_k^{\rm{H}}({\boldsymbol{q}},{{\boldsymbol{u}}_k})}{{\boldsymbol{\dot w}}_k}{\boldsymbol{\dot w}}_k^{\rm H}{{\boldsymbol{h}}_k({\boldsymbol{q}},{{\boldsymbol{u}}_k})}} \right)\nonumber \\
&=( {{{\boldsymbol{r}}^{\rm H}}{{{\boldsymbol{\dot W}}}_k}{\boldsymbol{r}}} )\left( {{{\boldsymbol{h}}_k^{\rm{H}}({\boldsymbol{q}},{{\boldsymbol{u}}_k})}{{{\boldsymbol{\dot W}}}_k}{{\boldsymbol{h}}_k({\boldsymbol{q}},{{\boldsymbol{u}}_k})}} \right)
\end{align}

Thus, we have
\begin{align} \label{W pos-semi}
{{\boldsymbol{r}}^{\rm H}}( {{{{\boldsymbol{\dot W}}}_k} - {{\bar {\boldsymbol{W}} }_k}} ){\boldsymbol{r}} \ge 0,\forall k \in \mathcal{K}. 
\end{align}
According to \eqref{W pos-semi}, we have ${{\boldsymbol{\dot W}}_k} - {{\boldsymbol{\bar W}}_k}\succeq 0$. Based on this fact together with ${{\boldsymbol{\dot R}}_s} \succeq 0$, it follows that ${{\boldsymbol{\bar R}}_s} = \sum\nolimits_{k = 1}^K {{{{\boldsymbol{\dot W}}}_k}}  + {{\boldsymbol{\dot R}}_s} - \sum\nolimits_{k = 1}^K {{{{\boldsymbol{\bar W}}}_k}}$ should be positive semidefinite.

Then, we show that the objective value achieved by $ \{ {{\boldsymbol{\bar W}}_k}\}$ and ${{\boldsymbol{\bar R}}_s}$ remains same as that by $\{ {{\boldsymbol{\dot W}}_k}\}$ and ${{\boldsymbol{\dot R}}_s} $. One can first derive that
\begin{align}
{{\boldsymbol{h}}_k^{\rm{H}}({\boldsymbol{q}},{{\boldsymbol{u}}_k})}{\bar {\boldsymbol{W}} _k}{{\boldsymbol{h}}_k({\boldsymbol{q}},{{\boldsymbol{u}}_k})}\!= \!{{\boldsymbol{h}}_k^{\rm{H}}({\boldsymbol{q}},{{\boldsymbol{u}}_k})}{\bar {\boldsymbol{w}} _k}\bar {\boldsymbol{w}} _k^H{{\boldsymbol{h}}_k({\boldsymbol{q}},{{\boldsymbol{u}}_k})} \!= \!{{\boldsymbol{h}}_k^{\rm{H}}({\boldsymbol{q}},{{\boldsymbol{u}}_k})}{{\boldsymbol{\dot W}}_k}{{\boldsymbol{h}}_k({\boldsymbol{q}},{{\boldsymbol{u}}_k})},\forall k\in \mathcal{K} . 
\end{align}
By substituting \eqref{r rankone} into the first term of $\bar r_k(\{{{\boldsymbol{W}}_k}\}, {\boldsymbol{R}}_s)$, we have
\begin{align}\label{r first}
\begin{array}{l}
{\mathop \sum \nolimits_{k = 1}^K } \left({\log _2}\left(\sum\nolimits_{i = 1}^K {{{\boldsymbol{h}}_k^{\rm{H}}({\boldsymbol{q}},{{\boldsymbol{u}}_k})}{{{\boldsymbol{\bar W}}}_i}{{\boldsymbol{h}}_k({\boldsymbol{q}},{{\boldsymbol{u}}_k})}}  + {{\boldsymbol{h}}_k^{\rm{H}}({\boldsymbol{q}},{{\boldsymbol{u}}_k})}{{{\boldsymbol{\bar R}}}_s}{{\boldsymbol{h}}_k({\boldsymbol{q}},{{\boldsymbol{u}}_k})} + {\sigma ^2}\right)\right)\\
=  {\mathop \sum \nolimits_{k = 1}^K } \left({\log _2}\left(\sum\nolimits_{i = 1}^K {{{\boldsymbol{h}}_k^{\rm{H}}({\boldsymbol{q}},{{\boldsymbol{u}}_k})}{{{\boldsymbol{\dot W}}}_i}{{\boldsymbol{h}}_k({\boldsymbol{q}},{{\boldsymbol{u}}_k})}}  + {{\boldsymbol{h}}_k^{\rm{H}}({\boldsymbol{q}},{{\boldsymbol{u}}_k})}{{{\boldsymbol{\dot R}}}_s}{{\boldsymbol{h}}_k({\boldsymbol{q}},{{\boldsymbol{u}}_k})} + {\sigma ^2}\right)\right).
\end{array}
\end{align}
For the second term of $\bar r_k(\{{{\boldsymbol{W}}_k}\}, {\boldsymbol{R}}_s)$, we have
\begin{align}\label{r second}
&\sum\nolimits_{k = 1}^K {\left(a_k + \sum\nolimits_{i=1,i \ne k}^K {{\rm{tr}}} ({\bf{B}}_k({{{\bf{\bar W}}}_i} - {\bf{W}}_i)) + {\rm{tr}}({\bf{B}}_k({{{\bf{\bar R}}}_s} - {\bf{R}}_s))\right)}\nonumber\\
&= \sum\limits_{k = 1}^K \left(a_k + {\sum\nolimits_i^K {  \frac{{{\log }_2}(e)}{2^{a_k}}{\bf{h}}_k^{\rm{H}}({\bf{q}},{{\bf{u}}_k})({{{\bf{\bar W}}}_i} - {\bf{W}}_i} ){{\bf{h}}_k}({\bf{q}},{{\bf{u}}_k})} \right.\nonumber\\
&\qquad \qquad \left. - {\frac{{{\log }_2}(e)}{2^{a_k}}{\bf{h}}_k^{\rm{H}}({\bf{q}},{{\bf{u}}_k})({{{\bf{\bar W}}}_k} - {\bf{W}}_k){{\bf{h}}_k}({\bf{q}},{{\bf{u}}_k})} + {\frac{{{\log }_2}(e)}{2^{a_k}}{\bf{h}}_k^{\rm{H}}({\bf{q}},{{\bf{u}}_k})({{{\bf{\bar R}}}_s} - {\bf{R}}_s){{\bf{h}}_k}({\bf{q}},{{\bf{u}}_k})}\right)\nonumber \\
 & = \sum\limits_{k = 1}^K \left(a_k +{\sum\nolimits_{i=1,i \ne k}^K { \frac{{{\log }_2}(e)}{2^{a_k}}{\bf{h}}_k^{\rm{H}}({\bf{q}},{{\bf{u}}_k})({{{\bf{\dot W}}}_i} - {\bf{W}}_i} ){{\bf{h}}_k}({\bf{q}},{{\bf{u}}_k})} \right. \nonumber\\
&\qquad\qquad\left. + \frac{{{\log }_2}(e)}{2^{a_k}}{{\bf{h}}_k^{\rm{H}}({\bf{q}},{{\bf{u}}_k})({{{\bf{\dot R}}}_s} - {\bf{R}}_s){{\bf{h}}_k}({\bf{q}},{{\bf{u}}_k})}\right).
\end{align}

By combining \eqref{r first} and \eqref{r second}, it holds that the objective value remains the same. By combining the facts above, it is verified that $\{ {{\boldsymbol{\bar W}}_k}\} $ and ${{\boldsymbol{\bar R}}_s}$ is also the optimal solution to (SDR5.$l$). This thus complete the proof.

\subsection{Detailed Derivation Procedure for \eqref{P7 objective trans}}
First, we denote ${\boldsymbol{A}}({\boldsymbol{q}}[n],{{\boldsymbol{u}}_k}) = {\boldsymbol{a}}({\boldsymbol{q}}[n],{{\boldsymbol{u}}_k}){{\boldsymbol{a}}^{\rm{H}}}({\boldsymbol{q}}[n],{{\boldsymbol{u}}_k})$. Accordingly, the objective function in (P7) is rewritten as
\begin{align}\label{23 first}
{\log _2}\left(\frac{{\sum\nolimits_{i = 1}^K {{\rm{tr}}\big({{\boldsymbol{W}}_i}[n]{\boldsymbol{A}}({\boldsymbol{q}}[n],{{\boldsymbol{u}}_k})\big)}  + {\rm{tr}}\big({{\boldsymbol{R}}_s}[n]{\boldsymbol{A}}({\boldsymbol{q}}[n],{{\boldsymbol{u}}_k})\big)\! +\! \frac{{{\sigma ^2}}}{\beta }({H^2}\! +\! \|{\boldsymbol{q}}[n] - {{\boldsymbol{u}}_k}\|{^2})}}{{\sum\nolimits_{i=1,i \ne k}^K {{\rm{tr}}\big({{\boldsymbol{W}}_i}[n]{\boldsymbol{A}}({\boldsymbol{q}}[n],{{\boldsymbol{u}}_k})\big)} \! + \!{\rm{tr}}\big({{\boldsymbol{R}}_s}[n]{\boldsymbol{A}}({\boldsymbol{q}}[n],{{\boldsymbol{u}}_k})\big) \!+\! \frac{{{\sigma ^2}}}{\beta }({H^2} + \|{\boldsymbol{q}}[n] - {{\boldsymbol{u}}_k}\|{^2})}}\right). 
\end{align}

Next, the entry in the $p$-th row and $q$-th column of ${\boldsymbol{A}}({\boldsymbol{q}}[n],{{\boldsymbol{u}}_k})$ is given as
\begin{align}\label{A}
\big[{{\boldsymbol{A}}({\boldsymbol{q}}[n],{{\boldsymbol{u}}_k})}\big]_{p,q} = {{\mathop{\rm e}\nolimits} ^{j2\pi \frac{d}{\lambda }(p- q)\frac{H}{{\sqrt {{H^2} + \|{\boldsymbol{q}}[n] - {{\boldsymbol{u}}_k}\|{^2}} }}}}.
\end{align}

It is observed from \eqref{A} that ${{\boldsymbol{W}}_i}[n]$ and ${{\boldsymbol{A}}({\boldsymbol{q}}[n],{{\boldsymbol{u}}_k})}$ are hermitian, and thus we have 
\begin{align}\label{23 second}
&{\rm tr}\big({{\boldsymbol{W}}_i}[n]{{\boldsymbol{A}}({\boldsymbol{q}}[n],{{\boldsymbol{u}}_k})}\big)\nonumber \\
&= \sum\limits_{p = 1}^M {\sum\limits_{q = 1}^M {\big[{{\boldsymbol W}_i}[n]\big]_{p,q}{{\rm{e}}^{j2\pi \frac{d}{\lambda }(q - p) \frac{H}{{{d}({\boldsymbol{q}}[n],{{\boldsymbol{u}}_k})}}}}} } \nonumber \\
&=\sum\limits_{\alpha {\rm{ = }}1}^M {\big[{{\boldsymbol W}_i}[n]\big]_{\alpha ,\alpha }}+ 2\sum\limits_{p = 1}^M {\sum\limits_{q = p + 1}^M {\big|\big[{{\boldsymbol W}_i}[n]\big]_{p,q}\big|{\rm{cos}}\left(\theta _{p,q}^{{\rm W}_i}[n]{\rm{ + }}2\pi \frac{d}{\lambda }(q - p)\frac{H}{{{d}({\boldsymbol{q}}[n],{{\boldsymbol{u}}_k})}}\right)} } .
\end{align}

By substituting \eqref{23 second} into \eqref{23 first}, \eqref{P7 objective trans} is finally obtained.

\end{appendix}

\end{document}